\documentclass{pasj01}

\Received{2019/01/26}
\Accepted{2019/09/10}

\usepackage{mathpazo}
\usepackage[T1]{fontenc} 

\newcommand{\eqb}{\begin{equation}}
\newcommand{\eqe}{\end{equation}}
\newcommand{\eqab}{\begin{eqnarray}}
\newcommand{\eqae}{\end{eqnarray}}

\newcommand{\defeq}{:=}

\newcommand{\pd}[2]{\frac{\partial #1}{\partial #2}}
\newcommand{\od}[2]{\frac{{\rm d} #1}{{\rm d} #2}}
\newcommand{\diff}[1]{{\rm d}#1}
\newcommand{\ls}[1]{_{\rm #1}}
\newcommand{\us}[1]{^{\rm #1}}

\newcommand{\eqref}[1]{(\ref{#1})}
\newcommand{\sgra}{Sgr~A$^\ast\,$}
\newcommand{\rcs}{\chi\ls{red}^2}
\newcommand{\rcsmin}{\chi\ls{red.min}^2}

\begin{document} 

\title{
A Significant Feature in the General Relativistic Time Evolution of the Redshift of Photons Coming from a Star Orbiting \sgra
}


\author{Hiromi \textsc{Saida}\altaffilmark{1},
\email{saida@daido-it.ac.jp}
Shogo \textsc{Nishiyama}\altaffilmark{2},
Takayuki \textsc{Ohgami}\altaffilmark{3},
Yohsuke \textsc{Takamori}\altaffilmark{4},
Masaaki \textsc{Takahashi}\altaffilmark{5},
Yosuke \textsc{Minowa}\altaffilmark{6},
Francisco \textsc{Najarro}\altaffilmark{7}, 
Satoshi \textsc{Hamano}\altaffilmark{8},
Masashi \textsc{Omiya}\altaffilmark{9},
Atsushi \textsc{Iwamatsu}\altaffilmark{10},
Mizuki \textsc{Takahashi}\altaffilmark{10},
Haruka \textsc{Gorin}\altaffilmark{2},
Tomohiro \textsc{Kara}\altaffilmark{2},
Akinori \textsc{Koyama}\altaffilmark{2},
Yosuke \textsc{Ohashi}\altaffilmark{2},
Motohide \textsc{Tamura}\altaffilmark{9, 11},
Schun \textsc{Nagatomo}\altaffilmark{12}, 
Tetsuya \textsc{Zenko}\altaffilmark{12},  and
Tetsuya \textsc{Nagata}\altaffilmark{12}
}

\altaffiltext{1}{Daido University, 
Naogya, Aichi 457-8530, Japan}
\altaffiltext{2}{Miyagi University of Education, 
Sendai, Miyagi 980-0845, Japan}
\altaffiltext{3}{Konan University, 
Kobe,  Hyogo 658-8501, Japan}
\altaffiltext{4}{National Institute of Technology, Wakayama College, 
Gobo, Wakayama 644-0023, Japan}
\altaffiltext{5}{Aichi University of Education, 
Kariya, Aichi 448-8542, Japan}
\altaffiltext{6}{Subaru Telescope, National Astronomical Observatory of Japan, Hilo, HI 96720, USA}
\altaffiltext{7}{Centro de Astrobiolog\'{i}a (CSIC/INTA), 28850 Madrid, Spain}
\altaffiltext{8}{National Astronomical Observatory of Japan, Tokyo 181-8588, Japan}
\altaffiltext{9}{Astrobiology Center, Tokyo, NINS, 181-8588, Japan}
\altaffiltext{10}{Tohoku University, Sendai, Miyagi 980-8578, Japan}
\altaffiltext{11}{The University of Tokyo, Tokyo, 113-0033, Japan}
\altaffiltext{12}{Kyoto University, Kyoto 606-8502, Japan}



\maketitle
\begin{abstract}
The star S0-2, orbiting the Galactic central massive black hole candidate \sgra, passed its pericenter in May 2018. 
This event is the first chance to detect the general relativistic (GR) effect of a massive black hole, free from non-gravitational physics. 
The observable GR evidence in the event is the difference between the GR redshift and the Newtonian redshift of photons coming from S0-2. 
Within the present observational precision, the 1st post-Newtonian (1PN) GR evidence is detectable. 
In this paper, we give a theoretical analysis of the time evolution of the 1PN GR evidence, under a presupposition that is different from used in previous papers. 
Our presupposition is that the GR/Newtonian redshift is always calculated with the parameter values (the mass of \sgra, the initial conditions of S0-2, and so on) determined by fitting the GR/Newtonian motion of S0-2 with the observational data. 
It is then revealed that the difference of the GR redshift and the Newtonian one shows two peaks before and after the pericenter passage. 
This double-peak-appearance is due to our presupposition, and reduces to a single peak if the same parameter values are used in both GR and Newtonian redshifts as considered in previous papers. 
In addition to this theoretical discussion, we report our observational data obtained with the Subaru telescope by 2018. 
The quality and the number of Subaru data in 2018 are not sufficient to confirm the detection of the double-peak-appearance. 
\end{abstract}

\section{Introduction}
\label{sec:intro}

The effects of general relativistic (GR) has already been distinguished observationally from non-GR effects, for example, in the following situations: the weak gravity in our solar system (e.g. \cite{ref:will2014}), the cosmic microwave background radiation (e.g. \cite{ref:hinshaw+2013} and \cite{ref:planck+2018}), and the gravitational waves radiated by stellar-size compact objects (e.g. \cite{ref:ligo+2016}). 
However, the GR effect of massive black holes (BHs) remains to be distinguished observationally from non-GR effects. 
A good probe of the quantitative assessment of GR effect of a massive BH is the star called S0-2 (in the Keck nomenclature) or S2 (in the very large telescope, VLT, nomenclature), that is orbiting \sgra (of mass $\approx 4\times 10^6 M_\odot$), with an orbital period of $\approx 16$ yr and a closest distance to \sgra of $\approx 100$ au. 
Because S0-2 is regarded as a test particle moving in the gravitational field of \sgra, the motion of S0-2 provides us with the pure GR effect free from non-gravitational physics \citep{ref:zucker+2006}. 
Measurements of the pure GR effect in the motion of S0-2 will enable us to test GR in the strong gravitational field of \sgra.\footnote{
From the results of the Planck satellite in 2018~\citep{ref:planck+2018}, a modified gravitational theory, such as the Starobinsky model, may be considered as a good candidate theory of gravity under some assumptions. 
However, such discussions are for the early/inflationary universe and seems not to be applicable to the Galactic center scale. 
Therefore, we assume GR that is to be compared with Newtonian gravity at Galactic center. 
A comment on the modified theories of gravity will be given at the end of section~\ref{sec:sd}. 
} 

Monitoring observations of the S0-2 motion can be performed by a few groups using large telescopes such as VLT, Keck, Gemini and Subaru. 
We have been monitoring the redshift of photons emitted by S0-2 from 2014 using Subaru \citep{ref:nishiyama+2018}, and American and European groups have been monitoring the position and redshift of S0-2 for about 20 years using other telescopes \citep{
ref:boehle+2016,
ref:gillessen+2017,
ref:parsa+2017,
ref:chu+2018,
ref:gravity+2018,
ref:gravity+2019,
ref:do+2019}. 
Until 2017, those observations had not revealed a clear deviation from the prediction of Newtonian gravity in the S0-2 motion. 
However, it has been expected that the deviation from the Newtonian prediction would become detectable in the redshift of photons coming from S0-2 during its pericenter passage in 2018 (e.g., \cite{ref:zucker+2006}). 
Recently, a detection of the combination of the special relativistic and gravitational Doppler effects has been reported by a European group~\citep{ref:gravity+2018,ref:gravity+2019} and by an American group~\citep{ref:do+2019}.

The evidence of GR being explored by using the large telescopes is theoretically expressed as the difference between the redshift predicted by GR and the one predicted by Newtonian gravity. 
Within the present observational precision, this GR evidence is detectable at the 1st post-Newtonian order. 
The redshift depends on some parameters, for example, the mass of \sgra and the initial conditions of the S0-2 motion. 
In this paper, we adopt the following presupposition on the treatment of the parameter values;

\begin{description}
\item[Presupposition:]
The GR redshift is always calculated with \emph{the best-fitting parameter values determined by fitting the GR motion} of S0-2 with the observational data. 
The Newtonian redshift is always calculated with \emph{the best-fitting parameter values determined by fitting the Newtonian motion} of S0-2 with the observational data.
\end{description}
The GR best-fitting values and the Newtonian ones are different. 
In order to confirm the validity of GR for the gravitational field of \sgra, it is useful to search for evidence of GR in the difference between the two best fits. 
In this paper, we report the time evolution of the difference between GR redshift and Newtonian redshift. 
Under our presupposition, it shows two peaks before and after the pericenter passage of S0-2. 
This ``double-peak-appearance'' has not been reported so far in the previous papers (e.g. \cite{ref:gravity+2019,ref:do+2019}). 
In the previous papers, the same parameter values, which have been carefully determined, have been used in both GR and Newtonian redshifts (see section~\ref{sec:theory.deviation}), and then the resultant single peak behavior has been discussed. 
If the GR is favored by two different approaches, such as the approach of the previous papers and the one under our presupposition, then the GR can be favored more definitely than the case using only one approach.

As a by-product of our presupposition, it is found that the statistical quantity $\rcs$, called the ``reduced-chi-squared'', is not useful for discriminating GR and Newtonian gravity within the present observational precision. 
Therefore, instead of the $\rcs$, we propose another quantity, denoted as $\delta z$ in this paper (section~\ref{sec:fitting.measure}), which expresses to what extent the double-peak-appearance determined by the observational data matches well with the theoretically expected form of the double-peak-appearance.
Furthermore, in this paper, we report our observational data obtained by the Subaru telescope in 2017 and 2018, together with the data already reported in our previous paper~\citep{ref:nishiyama+2018}. 
Due to bad weather conditions and instrumental instabilities, the quality and the number of the data in 2018 are not sufficient to confirm the detection of the double-peak-appearance, where the detection error is about $60\%$ according to our quantity $\delta z$. 
We need additional data sets to confirm the detection of the double-peak-appearance.

Section~\ref{sec:theory} is devoted to the theoretical discussion to derive the ``double-peak-appearance'' in the time evolution of the difference between the GR redshift and the Newtonian one under our presupposition. 
The non-usefulness of $\rcs$ for discriminating the GR and the Newtonian gravity within the present observational precision is also discussed. 
Section~\ref{sec:data} is the summary of our observations of S0-2 using the Subaru telescope from 2014 to 2018. 
In section~\ref{sec:fitting}, the best fit of the double-peak-appearance with our observational data is presented, and the quality of our 2018 data is also shown. 
Then, we introduce the quantity $\delta z$, which measures the discrepancy between the GR and the Newtonian gravity under our presupposition. 
Section~\ref{sec:sd} is the summary and discussion.

\section{Theoretically expected time evolution of the GR evidence under our presupposition}
\label{sec:theory}

\subsection{Definitions}
\label{sec:theory.definition}

The observational quantity that we focus on in this paper is the redshift $z$ of photons coming from S0-2 to the observer,
\eqb
\label{eq:theory.z}
 z(t) \defeq \frac{\nu\ls{S}(t+t\ls{R}(t))}{\nu\ls{O}(t)} - 1
 \,,
\eqe
where $t$ is the observation time, $\nu\ls{O}(t)$ is the frequency of photon at the observation, $\nu\ls{S}(t+t\ls{R}(t)\,)$ is the frequency of the observed photon when it was emitted by S0-2, and $t\ls{R}$ denotes the so-called Roemer time delay (i.e. the change of propagation time of a photon from S0-2 to the observer due to the motion of S0-2).\footnote{
In the format of Publication of the Astrophysical Society of Japan (PASJ), the parenthesis in $\nu\ls{S}(t+t\ls{R}(t))$ is replaced by the square brackets as $\nu\ls{S}[\,t+t\ls{R}(t)\,]$. 
Moreover, in appendix~\ref{app:approx}, the spacetime coordinates of a star $x^\mu(\tau) = (\,t(\tau)\,,\,r(\tau)\,,\,\theta(\tau)\,,\,\varphi(\tau)\,)$ is replaced as $x^\mu(\tau) = [\,t(\tau)\,,\,r(\tau)\,,\,\theta(\tau)\,,\,\varphi(\tau)\,]$. 
In PASJ, the double-usage of parenthesis seems to be forbidden even for mathematical symbols. 
Readers of the official printing version of this paper need to pay attention to such a condition.
}
We define the measure of the evidence of GR by
\eqb
\label{eq:theory.dz.def}
 \Delta z\ls{GR}(t) \defeq z\ls{GR}(t) - z\ls{NG}(t)
 \,,
\eqe
where $z\ls{GR}(t)$ is the redshift calculated by GR and $z\ls{NG}(t)$ is the redshift by Newtonian gravity (NG). 
These redshifts depend on certain parameters, such as the mass of \sgra and the initial conditions of S0-2 motion, which are explained explicitly later. 
Note that, throughout this paper, our presupposition on the treatment of the parameter values is that noted in section~\ref{sec:intro}.

The Newtonian redshift $z\ls{NG}(t)$ is exactly equal to the line-of-sight component of velocity calculated with Newtonian gravity,\footnote{
In astronomy, ``radial'' has the same meaning as ``line-of-sight''. 
However, we use the term ``line-of-sight'' velocity instead of ``radial'' velocity.
} 
\eqb
\label{eq:theory.zNG}
 z\ls{NG}(t) =
 \frac{1}{c}V\ls{S.NG\parallel}(t+t\ls{R}(t)) - \frac{1}{c}V\ls{O.NG\parallel}(t)
 \,,
\eqe
where $c$ is the light speed, and $V\ls{S.NG\parallel}$ and $V\ls{O.NG\parallel}$ are the line-of-sight velocity of, respectively, S0-2 and the observer whose positive direction is from the observer to S0-2. 
The velocity of S0-2 $\vec{V}\ls{S.NG}(t)$ is given by the Keplerian motion. 
Even when the velocity of observer $\vec{V}\ls{O.NG}$ is constant, its line-of-sight component $V\ls{O.NG\parallel}(t)$ depends on time due to the motion of S0-2.
The Roemer time delay in the Newtonian case is calculated by
\eqb
\label{eq:theory.tR}
 t\ls{R}(t) = 
 \frac{1}{c}\bigl|\vec{x}\ls{S}(t)-\vec{x}\ls{O}(t)\bigr|
 - \frac{1}{c}\bigl|\vec{x}\ls{S}(t\ls{ref})-\vec{x}\ls{O}(t\ls{ref})\bigr|
 \,,
\eqe
where $t\ls{ref}$ is the reference time when we set the delay zero, $\vec{x}\ls{O}(t)$ is the position of the observer at the observation time $t$, and $\vec{x}\ls{S}(t)$ is the position of S0-2 at which the observed photon (that is received by the observer at $t$) was emitted.\footnote{
In the Newtonian case, one may not include the Roemer time delay because the light speed is treated as infinity in the Newtonian dynamics. 
However, in this paper, we give priority to the fact that the light speed is finite, and introduce the Roemer time delay not only in the GR case but also in the Newtonian case.
} 
The time evolution of the position of S0-2, $\vec{x}\ls{S}(t)$, is determined by the Newtonian equations of motion.

The GR redshift $z\ls{GR}(t)$ is given from the GR definition of frequency,
\eqb
\label{eq:theory.nu}
 \nu\ls{S}(t) \defeq - K^\mu U_{{\rm S}\,\mu}\bigr|_t
 \quad,\quad
 \nu\ls{O}(t) \defeq - K^\mu U_{{\rm O}\,\mu}\bigr|_t
 \,,
\eqe
where $K^\mu$ is the four-wave-vector (tangent vector to null geodesic) of a photon coming from S0-2 to the observer, $U\ls{S}^\mu$ is the four-velocity (tangent vector to time-like geodesic) of S0-2, and $U\ls{O}^\mu$ is the four-velocity of the observer. 
We solve the geodesic equations in Hamilton's formalism. 
For example, the time-like geodesic equations for S0-2 are 
\eqb
\label{eq:theory.geodesic.full}
 \od{U_{{\rm S}\mu}(\tau)}{\tau} = - \pd{\mathcal{H}(U\ls{S},x\ls{S})}{x\ls{S}^\mu}
 \quad,\quad
 \od{x\ls{S}^\mu(\tau)}{\tau} = \pd{\mathcal{H}(U\ls{S},x\ls{S})}{U_{{\rm S}\mu}}
 \,,
\eqe
where $\tau$ is the affine parameter (the proper time) of S0-2, $x\ls{S}^\mu(\tau)$ is the spacetime position of S0-2, and the Hamiltonian is
\eqb
\label{eq:theory.H.full}
  \mathcal{H} \defeq \frac{1}{2} g^{\mu\nu}(x\ls{S}) U_{{\rm S}\mu} U_{{\rm S}\nu}
  \,,
\eqe
where $g^{\mu\nu}$ is the inverse of the metric tensor of Kerr spacetime. 
The null geodesic of photons and time-like geodesic of the observer are similarly formulated.
The Roemer time delay in the GR case, $t\ls{R}$, is given by a complicated combination of the solutions of all geodesic equations for S0-2, photon and observer. 
The exact definition of $t\ls{R}$ can be formulated, but we do not show it here because it is going to be approximated to the similar form with $t\ls{R}$ in equation~\eqref{eq:theory.tR} in the next subsection.

The set-up of the coordinate system has to be clarified. 
The detail of it is explained in appendix~\ref{app:coordinate}, and here let us summarize an important point: 
Our definitions of some quantities, for example the Roemer time delay, are not exactly the same as those used previous papers~\citep{ref:gravity+2018,ref:do+2019}. 
For example, we always take into account the finiteness of the distance between Sun and \sgra in calculating the Roemer time delay, while the time delay in the previous papers is approximated by the infinite distance limit. 
However, under the present observational uncertainties, such differences in the definitions of some quantities are not detectable.

\subsection{Post-Newtonian and post-Minkowskian approximations within observational precision}
\label{sec:theory.approximation}

Full GR formulation has a high numerical cost. 
In order to reduce the cost, we use the post-Newtonian (PN) and post-Minkowskian (PM) approximations (e.g. \cite{ref:poisson+2014}) of the S0-2 motion and photon propagation. 
Some numerical simulations for PN and PM approximations have been shown in \citet{ref:angelil+2010a} and \citet{ref:angelil+2010b}. 
However, without those simulations, we can justify the 1st order PN (1PN) approximation for the S0-2 motion and the 0th order PM (0PM) approximation for the photon propagation within the present observational precision.

The mass of \sgra and the orbital elements of S0-2 have already been estimated with a few \% 
uncertainties (\cite{ref:gravity+2018,ref:gravity+2019,ref:do+2019}). 
Using the mass of \sgra, $M\ls{SgrA} \approx 4\times 10^6 M_\odot$, and the pericenter distance of S0-2 to \sgra, $r\ls{peri} \approx 100$ au, we can evaluate the parameter for the PN expansion,
\eqb
\label{eq:theory.PNparameter}
 \varepsilon \approx \frac{2 G M\ls{SgrA}}{c^2 r\ls{peri}} \sim 10^{-3}
 \,,
\eqe
where $G$ is the Newton's constant. 
This gives the order of the 1PN term in the redshift $\approx c\, \varepsilon \sim 100$ km/s, and the 1.5th order PN (1.5PN) term $\approx c\, \varepsilon^{3/2} \sim 1$ km/s.  
On one hand, from all available observational data (by the end of 2018) of the redshift of American, European and our Japanese groups, the current averaged observational uncertainty of redshift is $\approx 38$ km/s. 
Therefore, the 1PN terms in $c\, z\ls{GR}$ (the components in $c\, z\ls{GR}$ depending not on the spin but on the mass of \sgra) is detectable, but 1.5PN (the largest component depending on the spin of \sgra) and higher order terms in $c\, z\ls{GR}$ are not.

Because the PN approximation is designed for a gravitationally bounded object like S0-2, the propagation of photons needs to be considered separately, for example, in the PM approximation. 
The 0th order PM (0PM) approximation corresponds to the photon propagating on the Minkowski spacetime with neglecting the effect of gravity. 
The peculiar effect in the 1st order PM (1PM) approximation is the gravitational lens effect. 
The bending angle $\delta\varphi$ of the photon orbit is estimated as
\eqb
 \delta\varphi \approx \frac{4GM\ls{SgrA}}{c^2\,r\ls{peri}}  \sim 10^{-3}
 \,.
\eqe
This is the same order as the PN parameter, $\delta\varphi \simeq \varepsilon$.
The relation between the propagation distance form S0-2 to the observer in the 1PM approximation, $L\ls{1PM}$, and the one in the 0PM approximation, $L\ls{0PM}$, is estimated as
\eqb
 L\ls{0PM} \simeq L\ls{1PM} \cos\delta\varphi
 \simeq L\ls{1PM} (1-\varepsilon^2)
 \,.
\eqe
Because the terms proportional to $\varepsilon^2$ is ignored within the present observational precision, it is enough for us to adopt the 0PM approximation of the photon propagation. 
The Roemer delay in the GR case with the 0PM approximation is also given by $t\ls{R}(t)$ in equation~\eqref{eq:theory.tR}, where the position of S0-2 in the 1PN case, $\vec{x}\ls{S.1PN}(t)$, is not necessarily equal to the one in the Newtonian case, $\vec{x}\ls{S.NG}(t)$, under our presupposition on the parameter values (see section~\ref{app:approx.analyses} of appendix~\ref{app:approx}).

For the motion of the observer, we can ignore the GR effect because of the huge distance to \sgra from us $\simeq 8$~kpc. 
We assume the velocity of the observer is constant.

The above discussions justify the 1PN approximation for the S0-2 motion and the 0PM approximation for the photon propagation. 
Hereafter, the combination of these approximations is phrased as the ``1PN+0PM'' approximation. 
Throughout this paper, the 1PN+0PM approximation is used under the assumption of the constant velocity of the observer.

The derivations of 1PN+0PM formulas are shown in appendix~\ref{app:approx}, and here we summarize them. 
The 1PN+0PM formula of the GR redshift can be expressed as
\eqab
 z\ls{1PN.0PM}(t) &=&
 \frac{1}{c} V\ls{S.1PN\parallel}(t+t\ls{R}) - \frac{1}{c}V\ls{O.1PN\parallel}(t)
 \nonumber
 \\
 &&+\frac{\vec{V}\ls{S.1PN}(t+t\ls{R})^2 - \vec{V}\ls{O}(t)^2}{2c^2}
 \nonumber
 \\
 && +\frac{G M\ls{SgrA}}{c^2\, r\ls{S.1PN}(t+t\ls{R})}
 \,,
\label{eq:theory.z1PN0PM}
\eqae
where $\vec{V}\ls{S.1PN}$ is the spatial velocity of S0-2 at the 1PN approximation, $\vec{V}\ls{O.1PN}$ is the constant velocity of the observer, $V\ls{S.1PN\parallel}$ and $V\ls{O.1PN\parallel}$ are the line-of-sight components of the velocities, and $r\ls{S.1PN}$ is the radial coordinate of S0-2 at the 1PN approximation. 
The second line in equation~\eqref{eq:theory.z1PN0PM}, $\vec{V}\ls{S.1PN}^2 - \vec{V}\ls{O.1PN}^2$, arises from the special relativistic Doppler effect at the 0PM approximation of the photon propagation. 
The third line in equation~\eqref{eq:theory.z1PN0PM}, $GM\ls{SgrA}/r\ls{S.1PN}$, arises from the gravitational Doppler effect at the 1PN approximation of the S0-2 motion. 
Note that the time evolutions of the velocity, $\vec{V}\ls{S.1PN}(t)$, and the radial coordinate, $r\ls{S.1PN}(t)$, are the solutions of the geodesic equations~\eqref{eq:theory.geodesic.full} with the 1PN Hamiltonian,
\eqab
 \mathcal{H}\ls{1PN} &=&
 -\Bigl( \frac{1}{2} + \frac{G M\ls{SgrA}}{c^2\,r\ls{S.1PN}}
           + \frac{2 G^2 M\ls{SgrA}^2}{c^4\,r\ls{S.1PN}^2}
   \Bigr) U_{{\rm S}t}^2
\nonumber
\\
 &&
 +\Bigl( \frac{1}{2} - \frac{G M\ls{SgrA}}{c^2\,r\ls{S.1PN}}
   \Bigr) U_{{\rm S}r}^2
\nonumber
\\
 &&
 +\frac{U_{{\rm S}\theta}^2}{2 r\ls{S.1PN}^2}
 +\frac{U_{{\rm S}\varphi}^2}{2 r\ls{S.1PN}^2 \sin^2\varphi\ls{S.1PN}}
 \,,
\label{eq:theory.H.1PN}
\eqae
where, because of the stationarity and axial symmetry of BH spacetime, the temporal and azimuthal components of the one-form $U_{{\rm S}\mu}$ are the constants of motion given by
\eqb
\label{eq:theory.constant}
 U_{{\rm S}t} = g_{t\mu}U\ls{S}^\mu = - \frac{E\ls{S}}{M\ls{S}c^2}
 \quad,\quad
 U_{{\rm S}\varphi} = g_{\varphi\mu}U\ls{S}^\mu = \frac{L\ls{S}}{M\ls{S}c}
 \,,
\eqe
where $E\ls{S}$, $L\ls{S}$ and $M\ls{S}$ are, respectively, the relativistic energy, angular momentum and the rest mass of S0-2. 
(In our numerical calculations, the values of the constants $U_{{\rm S}t}$ and $U_{{\rm S}\varphi}$ are determined by the initial conditions of the S0-2 motion, without specifying the values of $E\ls{S}$, $M\ls{S}$ and $L\ls{S}$.)
By solving the geodesic equations~\eqref{eq:theory.geodesic.full} with the 1PN Hamiltonian~\eqref{eq:theory.H.1PN}, we obtain the spacetime position $x\ls{S}^\mu(\tau)$ and the four-velocity $U\ls{S}^\mu(\tau) = g(x\ls{S}(\tau))^{\mu\nu}U_{{\rm S}\nu}(\tau)$ of S0-2 at the 1PN approximation. 
Note that, as shown by equation~\eqref{eq:approx.US} in appendix~\ref{app:approx}, we find for the spatial components of velocity, $U\ls{S}^i = V\ls{S.1PN}^i$ ($i = r,\,\theta,\,\varphi$), at the 1PN approximation.

Then, our measure of the GR evidence at the 1PN+0PM approximation is obtained by substituting equations~\eqref{eq:theory.zNG} and~\eqref{eq:theory.z1PN0PM} into equation~\eqref{eq:theory.dz.def},
\eqab
 \Delta z\ls{1PN.0PM}(t) &=&
 z\ls{1PN.0PM}(t) - z\ls{NG}(t)
\nonumber
\\
 &=&
 \frac{V\ls{S.1PN\parallel}(t+t\ls{R}) - V\ls{S.NG\parallel}(t+t\ls{R})}{c}
\nonumber
\\
 &&-\frac{V\ls{O.1PN\parallel}(t)-V\ls{O.NG\parallel}(t)}{c}
\nonumber
 \\
 &&+\frac{\vec{V}\ls{S.1PN}(t+t\ls{R})^2 - \vec{V}\ls{O.1PN}(t)^2}{2c^2}
\nonumber
\\
 &&+\frac{G M\ls{SgrA}}{c^2\, r\ls{S.1PN}(t+t\ls{R})}
 \,.
\label{eq:theory.dz1PN0PM}
\eqae
This $\Delta z\ls{1PN.0PM}(t)$ is the difference of the GR redshift and Newtonian redshift under our presupposition on the parameter values. 
Note that, while the Newtonian redshift $z\ls{NG}$ is given as an explicit function of the observation time $t$ by solving the Newtonian equations of motion, the GR redshift $z\ls{1PN.0PM}$ is, however, given as a function of the affine parameter $\tau$, not of $t$. 
The time $t$ in the GR redshift is, in its exact form, the coordinate time $t\ls{GR}(\tau)$ of the S0-2 motion given as a solution of geodesic equations. 
Therefore, we solve equation $t\ls{GR}(\tau) = t$ numerically for given $t$, when it is needed.

Some details on $\Delta z\ls{1PN.0PM}(t)$ are analyzed in section~\ref{app:approx.analyses} of appendix~\ref{app:approx}; here let us summarize an important point. 
The first and second terms in $\Delta z\ls{1PN.0PM}(t)$, which are the difference between the line-of-sight velocities of GR and Newtonian cases, do not necessarily vanish and have to be counted as the non-vanishing components in $\Delta z\ls{1PN.0PM}(t)$ under our presupposition. 
The reason is that the best-fitting values of parameters (e.g. the \sgra's mass and the S0-2's initial conditions) in the GR case is different from those in the Newtonian case, and hence the same quantities in both GR and Newtonian cases, such as the line-of-sight velocities of S0-2 and the observer, take different values in the GR and Newtonian cases.

\subsection{On the quantity that measures the difference between the GR and Newtonian predications}
\label{sec:theory.deviation}

In order to assess the deviation from the Newtonian prediction in the observational data of redshift $z\ls{obs}$, it is enough to calculate the difference,
\eqb
\label{eq:theory.dzobs}
  \Delta z\ls{obs} \defeq z\ls{obs} - z\ls{NG} \,,
\eqe
where $z\ls{NG}$ is the best-fitting Newtonian redshift. 
If $\Delta z\ls{obs}$ does not stay at zero for all observation times, then it is concluded that the observational data do not obey the Newtonian prediction. 
However, in order to assess not only the deviation from the Newtonian prediction but also the validity of GR, it is necessary to define a quantity to measure the evidence of GR. 
As such a quantity, we introduce the difference between the GR and Newtonian redshifts under our presupposition on the parameter values, $\Delta z\ls{1PN.0PM}(t)$ defined in equation~\eqref{eq:theory.dz1PN0PM}. 
\begin{itemize}
\item
Given the observational data, calculate $\Delta z\ls{1PN.0PM}(t)$ under our presupposition. 
Then, the closer the time evolution of $\Delta z\ls{1PN.0PM}(t)$ to the \emph{theoretically expected time evolution of it}, the more definite the detection of the difference between the GR and Newtonian predictions. 
\end{itemize}
Here, the point in this assessment is how we can estimate the theoretically expected time evolution of $\Delta z\ls{1PN.0PM}(t)$. 
The next subsection is devoted to this point.

In the previous papers~\citep{ref:gravity+2018,ref:gravity+2019,ref:gravity+2019,ref:do+2019}, the quantity to measure the GR evidence is different from our $\Delta z\ls{1PN.0PM}(t)$. 
The point of discussions in the previous papers is that their treatment of the parameter values is different from our presupposition. 
They have introduced an auxiliary parameter $f$ in the redshift formula as
\eqb
\label{eq:theory.zprev}
 z\ls{GR}\us{(prev)} =
 z\ls{NG}\us{(prev)}
 + f \times
 \left[\!\!\!
 \mbox{
 \begin{tabular}{l}
 2nd and 3rd lines
 \\
 in equation~\eqref{eq:theory.z1PN0PM}
 \end{tabular}
 } 
 \!\!\!\right]\us{(prev)}
\,,
\eqe
where the upper suffix ``(prev)'' denotes the treatment of parameter values in the previous papers. 
Their treatment is to determine all parameters including $f$ by fitting the observational data with the pure GR motion of S0-2 together with the modified redshift~\eqref{eq:theory.zprev}. 
They define the measure of GR evidence as
\eqab
\nonumber
 \Delta z\ls{GR}\us{(prev)}
 &\defeq&
 z\ls{GR}\us{(prev)} - z\ls{(NG)}\us{(prev)}
\\
\label{eq:theory.dzprev}
 &=&
 f \times
 \left[\!\!\!
 \mbox{
 \begin{tabular}{l}
 2nd and 3rd lines
 \\
 in equation~\eqref{eq:theory.z1PN0PM}
 \end{tabular}
 } 
 \!\!\!\right]\us{(prev)}
\,,
\eqae
where the same values of parameters are used in both terms $z\ls{GR}\us{(prev)}$ and $z\ls{NG}\us{(prev)}$. 
This $\Delta z\ls{GR}\us{(prev)}$ is called the ``GR effect'' in the previous papers. 
The points of this $\Delta z\ls{GR}\us{(prev)}$ can be summarized as follows:
\begin{itemize}
\item
In equation~\eqref{eq:theory.zprev}, the parameter $f$ is introduced by hand, while the geodesic equations of S0-2 (and of photons) are not modified by introducing the parameter $f$. 
This parameterization is different from the so-called Parametrized-Post-Newtonian (PPN) formalism, which are the parametrization of the spacetime metric tensor at the 1PN order and causes some modifications not only of the redshift of photons but also of the S0-2 motion. 
Because this $f$ is not exactly a parametrization used widely in the usual PPN formalism, the parameter $f$ is interpreted as an ad-hoc or a highly specialized parameter to measure the combination of the special relativistic and gravitational Doppler effects.
\item
Let the GR motion of S0-2 be substituted in $z\ls{GR}\us{(prev)}$. 
Then, the case of $f=1$ denotes the GR case, because $\Delta z\ls{GR}\us{(prev)}$ with $f=1$ is exactly the combination of the special relativistic and gravitational Doppler effects at the 1PN+0PM order. 
However, the case of $f=0$ never denotes a Newtonian case, because the ``GR motion'' of S0-2 is substituted in $z\ls{GR}\us{(prev)}$. 
In general, the case of $f\neq 1$ is not a modified theory of gravity, because the S0-2 motion is the pure GR case (i.e. the gravity is not modified for the S0-2 motion) while only the redshift formula is modified by introducing $f$. 
\item
From the above two points, the introduction of $f$ into $z\ls{GR}\us{(prev)}$ can be interpreted as the assessment of the hypothesis that the gravitational field of \sgra is described by GR (neither Newtonian gravity nor a modified theory of gravity). 
When the value of $f$ is determined by fitting the observational data with $z\ls{GR}\us{(prev)}$ together with the GR motion of S0-2, the closer the best-fitting value of $f$ to unity, the more plausible the hypothesis that the \sgra's gravity is GR. 
\end{itemize}
The quantity $\Delta z\ls{GR}\us{(prev)}$ is not a deviation from Newtonian prediction, but the measure to assess the ``GR hypothesis''. 
\citet{ref:gravity+2019} has reported the best-fitting value of $f = 1.04\pm 0.05$ using GRAVITY data by 2018, and \citet{ref:do+2019} has reported the best-fitting value of $f = 0.88\pm 0.16$ using Keck, Gemini and Subaru data by 2018. 
The evidence of GR has been found through the assessment of the GR hypothesis.

Both quantities $\Delta z\ls{1PN.0PM}$ and $\Delta z\ls{GR}\us{(prev)}$ can assess the validity of GR as the theory of gravity near \sgra, although the exact meanings of these quantities are different. 
Our quantity $\Delta z\ls{1PN.0PM}$ focuses on the total deviation of the GR prediction from the Newtonian prediction under our presupposition. 
The quantity in the previous papers $\Delta z\ls{GR}\us{(prev)}$ focuses on the combination of the special relativistic and gravitational Doppler effects, excluding the difference of the time evolution of S0-2's velocity between the GR and Newtonian cases. 
Note that, if the GR is favored by two different approaches, such as the approach of the previous papers and the one introduced in this paper, then the GR can be favored more definitely than the case favored by only one approach. 
Our approach does not conflict with the approach of the previous papers, 
but provides us with an additional reference for confirming the validity of GR. 
\footnote{
In the previous papers, in addition to the assessment of GR hypothesis, a direct comparison of the GR and Newtonian best-fitting orbits of S0-2 has been discussed through the Bayesian approach (e.g. using the so-called Bayes factor or Occam factor), or the so-called reduced-chi-squared $\rcs$ with putting higher weights on the data in 2018 than the other data. 
}

\subsection{Expected time evolution of $\Delta z\ls{1PN.0M}$ for ideally accurate observational data}
\label{sec:theory.evolution}

The theoretically expected time evolution of $\Delta z\ls{1PN.0PM}(t)$ is the key issue in this paper. 
Let us introduce a condition:
\begin{description}
\item[Condition (ideally accurate data set):]
An \emph{ideally accurate observational data set} is given. 
Here, the term ``ideally accurate'' denotes that (i) the error assigned to each data is constant for all observation times, and (ii) the observational value itself takes exactly the \emph{same value as the GR prediction}, where the offset of the astrometric origin is \emph{zero}. 
\end{description}
Under this condition, we define the theoretically expected time evolution of $\Delta z\ls{1PN0PM}(t)$ as the one derived by the following steps:
\begin{list}{}{}
\item[Step 1:]
Fix the values of all parameters which are listed in section~\ref{sec:theory.evolution.step1}. 
Using these values, calculate the GR motion of S0-2, $x\ls{S.1PN}^\mu(\tau)$, and the GR redshift, $z\ls{1PN.0PM}(t)$, at the 1PN+0PM approximation. 
\item[Step 2:]
\emph{Artificially create the ideally accurate data set}, in which every value of R.A., Dec. and redshift ($z\ls{1PN.0PM}$) of S0-2 are exactly the same with the GR prediction given in step~1, and the astrometric offset defined in equation~\eqref{eq:coordinate.Asky} is zero, $\vec{A}\ls{sky} \equiv \vec{0}$. 
Let the error assigned to each data be the averaged error of real observational data. 
\item[Step 3:]
By fitting the artificial data with the Newtonian motion of S0-2, calculate the Newtonian best-fitting values of all parameters listed in section~\ref{sec:theory.evolution.step1}. 
Such Newtonian best-fitting parameter values are not necessarily equal to the parameter values used in step~1. 
Then, calculate the Newtonian redshift, $z\ls{NG}(t)$, using the Newtonian best-fitting parameter values.
\item[Step 4:]
From the steps~1 and~3, calculate the time evolution of the quantity, 
$\Delta z\ls{1PN.0PM}(t) = z\ls{1PN.0PM}(t)-z\ls{NG}(t)$.
This is interpreted as the \emph{theoretically expected time evolution} of the difference between the GR and Newtonian redshifts under our presupposition on the parameter values.
\end{list}
In following subsections, we will carry out these steps.

\subsubsection{Step 1: Parameter values for GR prediction}
\label{sec:theory.evolution.step1}

\begin{longtable}{c||c|c|c|c|c|c}
\caption{
Two examples of parameter values with which GR motions are calculated. 
}
\label{tab:theory.paraGR}
\endfirsthead 
\hline
 Parameters for
 &
 $M\ls{SgrA}$ &
 $R\ls{GC}$ &
 $V\ls{O.ra}$ &
 $V\ls{O.dec}$ &
 $V\ls{O.Z}$ &
 ---
\\
 \sgra and observer
 &
 [$10^6 M_\odot$] &
 [kpc] &
 [mas/yr] &
 [mas/yr] &
 [km/s] &
 ---
\\
\hline
 \citet{ref:boehle+2016} &
 4.12 &
 8.02 &
 0.02 &
 $-$0.55 &
 $-$15 &
 ---
\\
\hline
 \citet{ref:gravity+2018} &
 4.100 &
 8.122 &
 $-$0.076 &
 $-$0.178 &
 1.9 &
 ---
\\
\hline\hline
 Parameters for
 &
 $I\ls{S}$ &
 $\Omega\ls{S}$ &
 $\omega\ls{S}$ &
 $e\ls{S}$ &
 $T\ls{S}$ &
 $t\ls{S.apo}$
\\
 S0-2 orbit
 &
 [deg] &
 [deg] &
 [deg] &
 [no dim.] &
 [yr] &
 [AD]
\\
\hline
 \citet{ref:boehle+2016} &
 134.7 &
 227.9 &
 66.5 &
 0.890 &
 15.90 &
 2010.293
\\
\hline
 \citet{ref:gravity+2018} &
 133.818 &
 227.85 &
 66.13 &
 0.88466 &
 16.0518 &
 2010.35384
\\
\hline
\end{longtable}

\begin{longtable}{c||c|c|c|c}
\caption{
Best fit of Newtonian motion of S0-2 with each data set created in step~2. 
$N$ is the number of data per year, for each of R.A., Dec. and $cz$. 
The error in $\chi^2$-fitting is given by definition~\eqref{eq:theory.formalerror}.
}
\label{tab:theory.paraNG}
\endfirsthead 
\hline
 $\rcsmin$ and parameters
 &
 $\rcsmin$ &
 $M\ls{SgrA}$ &
 $R\ls{GC}$ &
 $V\ls{O.ra}$
\\
 determined by $\chi^2$-fitting
 &
 [no dim.] &
 [$10^6 M_\odot$] &
 [kpc] &
 [mas/yr]
\\
\hline
 \authorcite{ref:boehle+2016}, $N=10$
 &
 0.0751 &
 $4.235\pm 0.028$ &
 $8.126\pm 0.026$ &
 $0.054\pm 0.002$
\\
\hline
 \authorcite{ref:boehle+2016}, $N=15$
 &
 0.0739 &
 $4.232\pm 0.023$ &
 $8.123\pm 0.021$ &
 $0.054\pm 0.002$
\\
\hline
 \authorcite{ref:boehle+2016}, $N=20$
 &
 0.0739 &
 $4.231\pm 0.020$ &
 $8.122\pm 0.018$ &
 $0.054\pm 0.002$
\\
\hline
 \authorcite{ref:gravity+2018}, $N=10$
 &
 0.0689 &
 $4.208\pm 0.027$ &
 $8.223\pm 0.025$ &
 $-0.045\pm 0.002$
\\
\hline
 \authorcite{ref:gravity+2018}, $N=15$
 &
 0.0687 &
 $4.205\pm 0.022$ &
 $8.220\pm 0.020$ &
 $-0.045\pm 0.002$
\\
\hline
 \authorcite{ref:gravity+2018}, $N=20$
 &
 0.0687 &
 $4.205\pm 0.019$ &
 $8.220\pm 0.018$ &
 $-0.045\pm 0.002$
\\
\hline\hline
Parameters
 &
 $V\ls{O.dec}$ &
 $V\ls{O.Z}$ &
 $I\ls{S}$ &
 $\Omega\ls{S}$
\\
 &
 [mas/yr] &
 [km/s] &
 [deg] &
 [deg]
\\
\hline
 \authorcite{ref:boehle+2016}, $N=10$
 &
 $-0.533\pm 0.002$ &
 $2.484\pm 1.512$ &
 $134.764\pm 0.084$ &
 $227.106\pm 0.096$
\\
\hline
 \authorcite{ref:boehle+2016}, $N=15$
 &
 $-0.533\pm 0.002$ &
 $2.422\pm 1.235$ &
 $134.756\pm 0.067$ &
 $227.113\pm 0.078$
\\
\hline
 \authorcite{ref:boehle+2016}, $N=20$
 &
 $-0.533\pm 0.001$ &
 $2.423\pm 1.070$ &
 $134.754\pm 0.058$ &
 $227.115\pm 0.067$
\\
\hline
 \authorcite{ref:gravity+2018}, $N=10$
 &
 $-0.163\pm 0.002$ &
 $19.029\pm 1.512$ &
 $133.872\pm 0.078$ &
 $227.111\pm 0.092$
\\
\hline
 \authorcite{ref:gravity+2018}, $N=15$
 &
 $-0.163\pm 0.002$ &
 $19.228\pm 1.235$ &
 $133.863\pm 0.064$ &
 $227.122\pm 0.075$
\\
\hline
 \authorcite{ref:gravity+2018}, $N=20$
 &
 $-0.163\pm 0.001$ &
 $19.248\pm 1.070$ &
 $133.863\pm 0.055$ &
 $227.122\pm 0.065$
\\
\hline\hline
Parameters
 &
 $\omega\ls{S}$ &
 $e\ls{S}$ &
 $T\ls{S}$ &
 $t\ls{S.apo}$
\\
 &
 [deg] &
 [no dim.] &
 [yr] &
 [AD]
\\
\hline
 \authorcite{ref:boehle+2016}, $N=10$
 &
 $65.817\pm 0.091$ &
 $0.8896\pm 0.0003$ &
 $15.8985\pm 0.0004$ &
 $2010.2956\pm 0.0005$
\\
\hline
 \authorcite{ref:boehle+2016}, $N=15$
 &
 $65.826\pm 0.073$ &
 $0.8896\pm 0.0002$ &
 $15.8985\pm 0.0003$ &
 $2010.2958\pm 0.0004$
 \\
\hline
 \authorcite{ref:boehle+2016}, $N=20$
 &
 $65.828\pm 0.063$ &
 $0.8896\pm 0.0002$ &
 $15.8985\pm 0.0003$ &
 $2010.2958\pm 0.0003$
\\
\hline
 \authorcite{ref:gravity+2018}, $N=10$
 &
 $65.497\pm 0.087$ &
 $0.8842\pm 0.0003$ &
 $16.0505\pm 0.0004$ &
 $2010.3566\pm 0.0005$
\\
\hline
 \authorcite{ref:gravity+2018}, $N=15$
 &
 $65.510\pm 0.071$ &
 $0.8842\pm 0.0002$ &
 $16.0503\pm 0.0003$ &
 $2010.3567\pm 0.0004$
 \\
\hline
 \authorcite{ref:gravity+2018}, $N=20$
 &
 $65.510\pm 0.061$ &
 $0.8842\pm 0.0002$ &
 $16.0503\pm 0.0003$ &
 $2010.3567\pm0.0004$
\\
\hline
\end{longtable}

As examples, let us use two sets of best-fitting parameter values given in \citet{ref:boehle+2016} and \citet{ref:gravity+2018}. 
Those values are shown in table~\ref{tab:theory.paraGR}, where the definitions of the eleven parameters are:\footnote{
In the published version in the journal (Publications of the Astrophysical Society of Japan), the horizontal lines in table~\ref{tab:theory.paraGR} are removed except for the line between the rows of ``Parameters for \sgra and observer'' and ``\citet{ref:boehle+2016}''. 
Therefore, the distinction between the upper list (from $M\ls{SgrA}$ to $V\ls{O.zZ}$) and the lower list (from $T\ls{S}$ to $t\ls{S.apo}$) disappears in the official printing version of this paper. 
The same problem is applied to all tables. 
Readers of the official printing version of this paper need to care about such a table style. 
}
\begin{list}{}{}
\item[$M\ls{SgrA}$:]
the mass of \sgra.
\item[$R\ls{GC}$:]
the distance between Sun and \sgra.
\item[$V\ls{O.ra}$:]
the $Y$ (R.A.)-component of the observer's velocity $\vec{V}\ls{O}$ relative to \sgra, see equation~\eqref{eq:coordinate.PS}.
\item[$V\ls{O.dec}$:]
the $X$ (Dec.)-component of the observer's velocity $\vec{V}\ls{O}$ relative to \sgra, see equation~\eqref{eq:coordinate.PS}.
\item[$V\ls{O.Z}$:]
the $Z$-component of the observer's velocity $\vec{V}\ls{O}$ relative to \sgra, see equation~\eqref{eq:coordinate.PS}.
\item[$I\ls{S}$:]
the inclination angle of the orbital plane of S0-2, when it is evaluated in the Newtonian motion.
\item[$\Omega\ls{S}$:]
the angle of ascending node from Dec. direction on the orbital plane of S0-2, when it is evaluated in the Newtonian motion.
\item[$\omega\ls{S}$:]
the angle of pericenter node from the ascending node on the orbital plane of S0-2, when it is evaluated in the Newtonian motion.
\item[$e\ls{S}$:]
the eccentricity of the S0-2 orbit, when it is evaluated in the Newtonian motion.
\item[$T\ls{S}$:]
the orbital period of S0-2 around \sgra, when it is evaluated in the Newtonian motion.
\item[$t\ls{S.apo}$:]
the time of the previous apocenter passage in 2010.
\end{list}
Here we need to note two remarks. 
The first remark is on the artificial data that will be created in step~2. 
We define the artificial data as the \emph{ideally accurate data} in which the astrometric offset defined in equation~\eqref{eq:coordinate.Asky} is not introduced, $\vec{A}\ls{sky}(t) \equiv \vec{0}$. 
Therefore, in table~\ref{tab:theory.paraGR}, the parameters corresponding to $\vec{A}\ls{sky}(t)$ are not included.

The second remark is on the last six parameters, from $I\ls{S}$ to $t\ls{S.apo}$. 
Although these six parameters are given in the form of orbital elements of the Newtonian motion, it does never mean that these six parameters are available only for the Newtonian motion. 
In solving the geodesic equations~\eqref{eq:theory.geodesic.full} of the S0-2 motion, we simply transform those six parameters to the initial conditions, position and velocity, given at the time $t\ls{S.apo}$. 
We regard those six parameters, from $I\ls{S}$ to $t\ls{S.apo}$, as the control parameters of the initial conditions for the GR motion. 
Hence, if the GR motion is given (for example, from the best-fitting calculation), then the position and velocity of S0-2 at the apocenter are transformed to the six parameters, $I\ls{S}$ to $t\ls{S.apo}$ by simple Newtonian formulas of these six parameters.

\subsubsection{Step 2: Ideally accurate data set}
\label{sec:theory.evolution.step2}

For each set of parameter values in table~\ref{tab:theory.paraGR}, we create the ideally accurate data set under the following conditions:
\begin{list}{}{}
\item[Condition 1:]
Create $N$ data of R.A., Dec. and $cz\ls{1PN.0PM}$ per year with a constant temporal interval, $1/N$ yr. 
\item[Condition 2:]
Create the data set corresponding to $L$ years' observations, where $L$ is sufficiently longer than one period, $T\ls{S}$, in order to follow the whole time evolution of $z\ls{GR}(t)$ in one period.\footnote{
We are interested in the physical property of $\Delta z\ls{GR}$, which appears in the time evolution within a period $\approx T\ls{S}$. 
Hence, we make the ideally accurate data set cover at least one period of the S0-2 motion. 
} 
This $L$ needs to be short enough to make the shift of the pericenter/apocenter angle be significantly smaller than $90^\circ$, because a large shift of the angle causes a significant change in the observed time evolution of $z\ls{GR}(t)$. 
\item[Condition 3:]
As noted at the beginning of this section~\ref{sec:theory.evolution}, the error assigned to each data is the averaged error of real observational data. 
The error in R.A. observation is $9.832\times 10^{-4}$ arcsec, in Dec. observation $9.176\times 10^{-4}$, and in redshift (times the light speed, $c z\ls{GR}$) observation $38.29$ km/s, which are read from the public data in \citet{ref:boehle+2016}, \citet{ref:gravity+2018} and our observations listed in table~\ref{tab:subaruUncertainty}.
\end{list}
The number of each kind of data, R.A., Dec. and $cz\ls{1PN.0PM}$, is $NL$ (i.e. $3NL$ data in total). 
In the following numerical calculations, we set $L=4 T\ls{S} \approx 64$~yr, during an interval $t\ls{S.apo}-2T\ls{S} < t < t\ls{S.apo}+2T\ls{S}$, centered at the previous apocenter time in 2010. 
This duration of $4T\ls{S}$ corresponds to the angle of pericenter/apocenter shift $\approx 4\times 6\pi GM\ls{SgrA}/(c^2 r\ls{S}) \sim 4^\circ$, which is sufficiently smaller than $90^\circ$. 
Further, we consider three cases of the number of data per year, $N = 10,\,15,\,20$, where $N=15$ is roughly the averaged number of real observations per year until 2017.  
Because we consider three values of $N$ for each example of parameters in table~\ref{tab:theory.paraGR}, we have six cases of artificial data sets. 
For these cases, we are going to calculate the expected time evolution of $\Delta z\ls{1PN0PM}$ under our presupposition on the parameter values. 
Our numerical calculations are performed using Mathematica, version~11.

\subsubsection{Step 3:  Fitting with Newtonian prediction}
\label{sec:theory.evolution.step3}

We carry out the $\chi^2$-fitting of the S0-2 Newtonian motion with each artificial data set created in step~2. 
The fitting method is a simple minimum search of the reduced-chi-squared, $\rcs$, and we have stopped the minimum search when the improvement of $\rcs$ becomes less than $10^{-4}$. 
As the initial-guess values of the parameters in the $\chi^2$-fitting, it is good to use the parameter values used in creating the ideally accurate data set, because the resultant minimum value of reduced-chi-squared, $\rcsmin$, of various initial-guesses coincide with each other within differences less than $10^{-2}$. 
Then, the best-fitting values of the parameters for every six data set created in step~2 are shown in table~\ref{tab:theory.paraNG}. 
The fitting error $\delta X$ of parameter $X = M\ls{SgrA}, R\ls{GC}, \cdots, t\ls{S.apo}$ in table~\ref{tab:theory.paraNG} is the formal error defined by~\citep{ref:press+1992}
\eqb
\label{eq:theory.formalerror}
 \delta X \defeq \sqrt{C_{XX}} \,,
\eqe
where $C$ is the covariance matrix (the inverse of the Hessian of ``chi-squared'' $\chi^2$ times $1/2$), and $C_{XX}$ is the diagonal element corresponding to the parameter $X$. 
This error~\eqref{eq:theory.formalerror} corresponds to $1\,\sigma$ error in the $\chi^2$-fitting when one parameter $X$ is varied and if each observational data has perfectly obeyed a Gaussian probabilistic distribution.

Before proceeding to step~4, let us remark an implication by the very small value of $\rcsmin \approx 0.07$ in table~\ref{tab:theory.paraNG}. 
Because table~\ref{tab:theory.paraNG} is made from the ideally accurate data sets, the $\rcsmin$ in table~\ref{tab:theory.paraNG} can be interpreted as one quantity that measures a discrepancy between GR and Newtonian gravity under the idea of $\chi^2$-assessment.\footnote{
Note that the $\chi^2$-assessment for discriminating some theories and the $\chi^2$-fitting for searching the best-fitting parameter values of each theory are different. 
In this paragraph we discuss only on the $\chi^2$-assessment.
} 
Therefore, if $\rcsmin$ in table~\ref{tab:theory.paraNG} was of the order of one or more, $\rcsmin \gtrsim O(1)$, then it was expected that we would be able to confirm the detection of $\Delta z\ls{1PN.0PM}(t)$ by the $\chi^2$-assessment. 
In other words, the small value $\rcsmin \approx 0.07$ in table~\ref{tab:theory.paraNG} implies that the $\chi^2$-assessment does not work well for a detection of $\Delta z\ls{1PN.0PM}(t)$, even if very accurate observations would be performed with the present observational precision.

\subsubsection{Step 4: Theoretically expected time evolution of our GR evidence $\Delta z\ls{1PN.0PM}(t)$}
\label{sec:theory.evolution.step4}

The parameter values in tables~\ref{tab:theory.paraGR} and~\ref{tab:theory.paraNG} provide us with a theoretically expected time evolution of $\Delta z\ls{1PN.0PM}(t)$. 
In this paper, all figures of redshift are shown in the unit of km/s, by multiplying the light speed as $cz(t)$.

\begin{figure}
 \begin{center}
 \includegraphics[width=80mm]{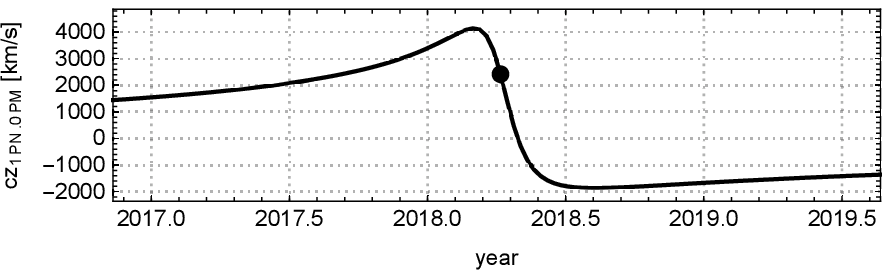} 
\\
 \includegraphics[width=80mm]{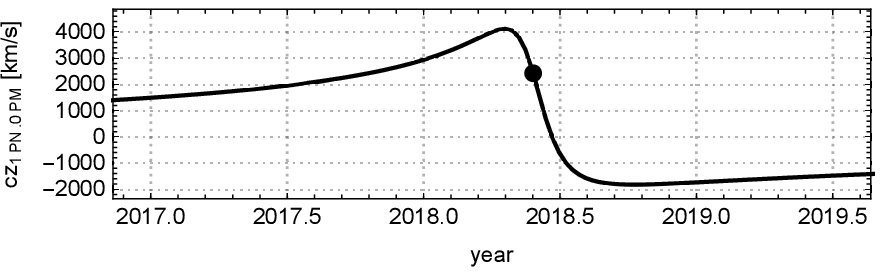}
  \end{center}
\caption{Time evolution of the observed redshift, $cz\ls{1PN.0PM}(t)$. 
Top panel is for the parameter values in \citet{ref:boehle+2016}. 
The bottom panel is for the parameter values in \citet{ref:gravity+2018}. 
Dots on the curves denote the pericenter passage of S0-2. 
}
\label{fig:theory.z1PN0PM}
\end{figure}

\begin{figure}
 \begin{center}
 \includegraphics[width=80mm]{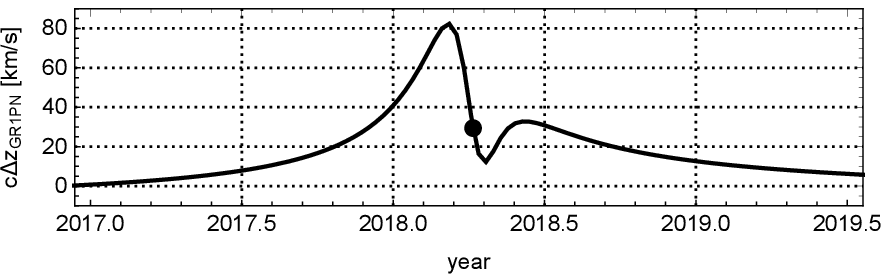} 
\\
 \includegraphics[width=80mm]{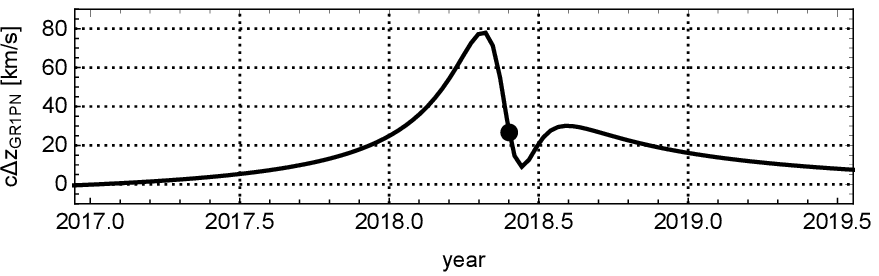}
  \end{center}
\caption{Theoretically expected time evolution of $c \Delta z\ls{1PN.0PM}(t)$. 
Top panel is for the case of \authorcite{ref:boehle+2016} with $N=20$ in table~\ref{tab:theory.paraNG}.
The bottom panel is for the case of \authorcite{ref:gravity+2018} with $N=10$ in table~\ref{tab:theory.paraNG}.
Dots on the curves denote the pericenter passage of S0-2 estimated by the 1PN+0PM approximation. 
All other cases in table~\ref{tab:theory.paraNG} show almost the same behavior. 
}
\label{fig:theory.GReffect}
\end{figure}

Figure~\ref{fig:theory.z1PN0PM} shows the theoretically expected time evolution of the redshift of photons coming from S0-2 at the 1PN+0PM approximation, $cz\ls{1PN.0PM}(t)$ in equation~\eqref{eq:theory.z1PN0PM}, using the parameter values in table~\ref{tab:theory.paraGR}. 
The top panel is for the case of \citet{ref:boehle+2016}, and the bottom panel for the case of \citet{ref:gravity+2018}. 
Hereafter, the dots attached on curves in the figures denote the pericenter and apocenter passages of S0-2 estimated by the 1PN+0PM approximation, not by the Newtonian case. 
The pericenter time in the Newtonian case is delayed slightly by $0.002$~yr $\approx 0.7$~day after the pericenter time in the 1PN+0PM approximation.\footnote{
The gravitational potential of BH estimated in GR is stronger than the one in Newtonian gravity. 
This makes the speed of S0-2 in the GR case tend to be greater than the speed in the Newtonian case, and the pericenter time in the GR case precedes the pericenter time in the Newtonian case.
} 
Note that, because the parameter values in \citet{ref:gravity+2018} are based on observations until June 2018 while those in \citet{ref:boehle+2016} are based on observations until 2013, we find a horizontal shift between the top and bottom panels in figure~\ref{fig:theory.z1PN0PM}. 
The discrepancy probably arises from the five-year difference of the observations. 
However, this discrepancy does not affect the result of this section.

Under our presupposition on the treatment of the parameter values, figure~\ref{fig:theory.GReffect} shows the theoretically expected time evolution of $c \Delta z\ls{1PN.0PM}(t)$. 
The upper and bottom panels correspond, respectively, to the cases of \authorcite{ref:boehle+2016} with $N=20$ and \authorcite{ref:gravity+2018} with $N=10$ in table~\ref{tab:theory.paraNG}. 
It is significant that the \emph{two peaks appear before and after the pericenter passage} in both panels. 
Although a horizontal shift is recognized between the two panels, as already seen in figure~\ref{fig:theory.z1PN0PM}, the ``double-peak-appearance'' is not affected by the horizontal shift between the two panels. 
The time evolution of $c \Delta z\ls{1PN.0PM}(t)$ for the other set of parameters in table~\ref{tab:theory.paraNG} also show the very similar ``double-peak-appearance'', although that is not presented here.

In order to understand the origin of the ``double-peak-appearance'', it is useful to consider each component of $c \Delta z\ls{1PN.0PM}(t)$ defined in equation~\eqref{eq:theory.dz1PN0PM}. 
As an example, we focus on the case of \authorcite{ref:gravity+2018} with $N=10$ in table~\ref{tab:theory.paraNG}.

\begin{figure}
 \begin{center}
 \includegraphics[width=80mm]{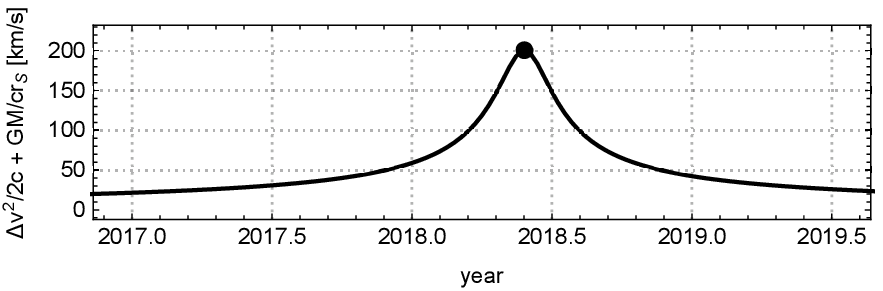} 
\\
 \includegraphics[width=77mm]{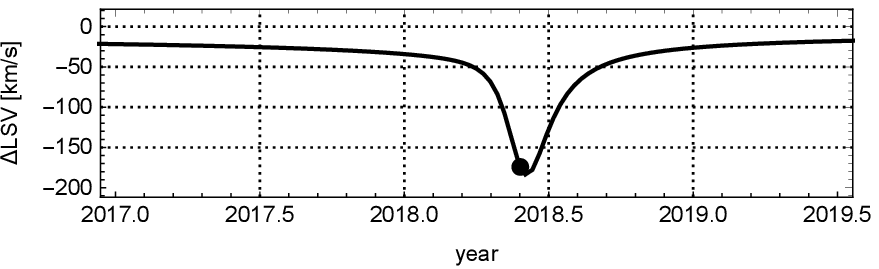} 
 \end{center}
\caption{
Decomposition of $\Delta z\ls{1PN.0PM}$ into two components. 
The top panel is for the theoretically expected time evolution of the sum of the third and fourth terms in equation~\eqref{eq:theory.dz1PN0PM}, $\Delta v\ls{S}^2/(2c)+GM\ls{SgrA}/(cr\ls{S})$. 
The bottom panel is for the theoretically expected time evolution of the sum of the first and second terms in equation~\eqref{eq:theory.dz1PN0PM} (noted as $\Delta$LSV in the panel). 
Both panels are drown with the case of \authorcite{ref:gravity+2018} with $N=10$. 
Dots on the curves denote the pericenter passage of S0-2 estimated by the 1PN+0PM approximation. 
}
\label{fig:theory.GReffect.component}
\end{figure}

The top panel in figure~\ref{fig:theory.GReffect.component} shows the theoretically expected time evolution of the sum of the third and fourth terms of $c \Delta z\ls{1PN.0PM}(t)$ in equation~\eqref{eq:theory.dz1PN0PM},
\eqb
\label{eq:theory.SRGRterm}
 \frac{\vec{V}\ls{S.1PN}(t+t\ls{R})^2 - \vec{V}\ls{O.1PN}(t)^2}{2c}
 +\frac{G M\ls{SgrA}}{c\, r\ls{S.1PN}(t+t\ls{R})}
 \,.
\eqe
This summation is the ``special relativistic and gravitational Doppler'' component in $c \Delta z\ls{1PN.0PM}(t)$ that has already been recognized in the previous papers. 
On the other hand, the bottom panel in figure~\ref{fig:theory.GReffect.component} shows the theoretically expected time evolution of the sum of the first and second terms of $c \Delta z\ls{1PN.0PM}(t)$ in equation~\eqref{eq:theory.dz1PN0PM},
\eqab
  \bigl[\, V\ls{S.1PN\parallel}(t+t\ls{R}) &-& V\ls{S.NG\parallel}(t+t\ls{R})\,\bigr]
\nonumber
 \\
 &-& \big[\, V\ls{O.1PN\parallel}(t)-V\ls{O.NG\parallel}(t) \,\bigr]
 \,.
\label{eq:theory.LSVterm}
\eqae
This summation is the ``line-of-sight velocity (LSV)'' component in $c \Delta z\ls{1PN.0PM}(t)$, and has not been considered so far in the previous papers. 
Note that tables~\ref{tab:theory.paraGR} and~\ref{tab:theory.paraNG} imply that the difference of observer's LSV is estimated to be $V\ls{O.1PN\parallel}(t)-V\ls{O.NG\parallel}(t) \approx V\ls{O.Z(1PN)}-V\ls{O.Z(NG)} \simeq -17$ km/s. 
This is smaller by one order than the LSV component in the bottom panel of figure~\ref{fig:theory.GReffect.component} $\approx -200$ km/s. 
Therefore, the time evolution of LSV component is determined mainly by the LSV of S0-2, $V\ls{S.1PN\parallel}(t+t\ls{R}) - V\ls{S.NG\parallel}(t+t\ls{R})$. 
Some theoretical analyses on this LSV component are given in the section~\ref{app:approx.analyses} of appendix~\ref{app:approx}.

The point in the LSV component~\eqref{eq:theory.LSVterm} is that, as indicated by the bottom panel of figure~\ref{fig:theory.GReffect.component}, the LSV of S0-2 in the Newtonian best-fitting case becomes faster than the LSV in the GR case, $V\ls{S.NG\parallel} > V\ls{S.1PN\parallel}$, around the pericenter passage. 
This is reasonable due to the following facts:
\begin{list}{}{}
\item[(i)]
In general, the $\chi^2$-fitting provides us with the parameter values that minimize the discrepancy between theory and data. 
Therefore, all sets of parameter values in table~\ref{tab:theory.paraNG} must be adjusted so that the orbit and redshift of S0-2 in the Newtonian case become as similar as possible to those in the GR case.
\item[(ii)]
The Newtonian redshift, $cz\ls{NG}(t)$ in equation~\eqref{eq:theory.zNG}, includes no counter-term to the ``special relativistic and gravitational Doppler'' component~\eqref{eq:theory.SRGRterm}. 
\end{list}
By facts (i) and (ii), it is expected that the motion of S0-2 with the Newtonian best-fitting parameter values is adjusted so as to compensate the special relativistic and gravitational Doppler component~\eqref{eq:theory.SRGRterm}. 
Further, because of fact (ii), it is only the LSV component $V\ls{S.NG\parallel}(t+t\ls{R})$ in the Newtonian motion of S0-2 that can compensate the special relativistic and gravitational Doppler component. 
Hence, as shown in figure~\ref{fig:theory.GReffect.component}, the LSV component~\eqref{eq:theory.LSVterm} takes the negative value $\approx -200$~km/s (bottom panel of figure~\ref{fig:theory.GReffect.component}) so as to compensate the positive value $\approx 200$~km/s of the special relativistic and gravitational Doppler component (top panel of figure~\ref{fig:theory.GReffect.component}). 
This means that the LSV of the Newtonian best-fit is faster than the LSV of the GR case.

From the above discussions, we find that, under our presupposition on the parameter values, the time evolution of $c\Delta z\ls{1PN.0PM}(t)$ shows the ``double-peak-appearance'' as in figure~\ref{fig:theory.GReffect}. 
In contrast with our presupposition, if one uses the method of the other groups summarized in section~\ref{sec:theory.deviation}, their quantity $\Delta z\ls{GR}\us{(prev)}$ defined in equation~\eqref{eq:theory.dzprev} shows a single peak feature similar to the one in the top panel of figure~\ref{fig:theory.GReffect.component}. 
(Note that the top panel of figure~\ref{fig:theory.GReffect.component} corresponds to the case $f=1$ of $\Delta z\ls{GR}\us{(prev)}$.)

\begin{figure}
 \begin{center}
 \includegraphics[width=80mm]{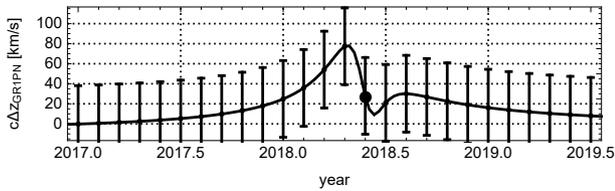} 
  \end{center}
\caption{Time evolution of $\Delta z\ls{1PN.0PM}(t)$ plotted with the artificial accurate data set created in step~2, for the case of \authorcite{ref:gravity+2018} with $N=10$. 
}
\label{fig:theory.GReffect.ArtData}
\end{figure}

\begin{figure}
 \begin{center}
 \includegraphics[width=80mm]{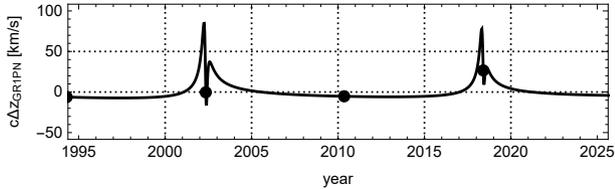} 
  \end{center}
\caption{Time evolution of $\Delta z\ls{1PN.0PM}(t)$ from 2000 to 2020, for the case of \authorcite{ref:gravity+2018} with $N=10$. 
Dots denote the pericenter and apocenter passage of S0-2 estimated by 1PN+0PM approximation. 
All other cases in table~\ref{tab:theory.paraNG} show almost the same behavior. 
}
\label{fig:theory.GReffect.Wide}
\end{figure}

Finally in this section, we show $c\Delta z\ls{1PN.0PM}(t)$ together with the artificial data in figure~\ref{fig:theory.GReffect.ArtData} for the case of \authorcite{ref:gravity+2018} with $N=10$. 
Further, because the existing real observational data of S0-2 covers the previous pericenter passage in 2002, we show in figure~\ref{fig:theory.GReffect.Wide} the theoretically expected time evolution of $c\Delta z\ls{1PN.0PM}(t)$ for a rather wide temporal range. 
As implied by this figure, the theoretically expected time evolution of $c\Delta z\ls{1PN.0PM}(t)$ under our presupposition shows the ``double-peak-appearance'' not only for the pericenter passage in 2018 but also for that in 2002.

\section{Our observations and data analysis}
\label{sec:data}

Readers who want to see the results of the fitting of our observational data and the ``double-peak-appearance'' may refer to our observational data in table~\ref{tab:subaruUncertainty} and go to section~\ref{sec:fitting}.

The observational data used in our fitting calculation are all public data released by 2017 \citep{ref:boehle+2016, ref:gillessen+2017} and our spectroscopic data obtained with the Subaru telescope by 2018. 
We have observed S0-2 for more than 10 nights with the Subaru telescope. 
However, due to unfortunate bad weather conditions at Hawaii island in 2018, 
we have obtained spectra with lower SN ratios than the previous years.
Our spectroscopic data, including ones reported in our previous paper~\citep{ref:nishiyama+2018}, are listed in table~\ref{tab:subaruUncertainty}. 
Details on our Subaru observations are as follows.

\subsection{Observation}
\label{subsec:obs}

We have carried out spectroscopic observations of S0-2 using the Subaru telescope \citep{Iye04Subaru} and IRCS \citep{00KobayashiSPIE}, in the Echelle mode.
The spectral resolution of the IRCS Echelle mode is $\approx 20,000$ in the $K$ band. 
During our observations, we have used the Subaru AO system \citep{Hayano08SPIE, Hayano10SPIE} and the laser guide star (LGS) system \citep{12MinowaSPIE}.
In the LGS mode observations, $R = 13.8$\,mag star USNO 0600-28577051 was used as a tip-tilt guide star, and in the natural guide star (NGS) mode, the star was used as the NGS. 
The details of the observations, such as exposure time and the number of frames taken in the nights, are shown in table \ref{Tab:Obs}.
The details of the observations from 2014 to 2016 are also described in \citet{ref:nishiyama+2018}.

\subsection{Data reduction}
\label{subsec:Reduction}

The reduction procedure for our data sets includes:
(1)~dark subtraction; 
(2)~flat-fielding;
(3)~sky subtraction;
(4)~bad pixel correction; and
(5)~cosmic ray removal.
A sky field was observed once or twice per night, and used for the correction of atmospheric emission. 
The S0-2 spectra are then extracted from the reduced images. 
The wavelength calibration was carried out using the sky OH emission lines. 
Spectra of nearby early-A type stars was used for the telluric correction. 
The details of the procedure above are described in \citet{ref:nishiyama+2018}.

\subsection{Combining the S0-2 Spectra}
\label{subsec:S0-2spec}

To determine the profile of the Br-$\gamma$ absorption line and redshifts of S0-2 accurately, we have combined spectra of S0-2 from 2014 to 2017. 
In our previous paper \citep{ref:nishiyama+2018}, we fitted the Br-$\gamma$ line using a Moffat function with all parameters set as free. 
However, since some low signal-to-noise (SN) ratio spectra are included in our data sets, the line shape could be different in such low SN ratio spectra. 
Hence we have combined S0-2 spectra from 2014 May to 2017 Aug, to determine the profile of the Br-$\gamma$ absorption line with a good SN ratio. 
Here we have not combined the spectra in 2018, because the redshift of S0-2 changes rapidly hour by hour. 

To combine S0-2 spectra, first we fit the Br-$\gamma$ line in each spectrum from 2014 to 2017 with a Moffat function, and determine the peak wavelength. 
The spectra are shifted to have zero redshift using IRAF {\it dopcor} task, and then combined to make a preliminary combined spectrum. 
Next, the Br-$\gamma$ line in the preliminary spectrum is fitted to determine the parameter of the Moffat function. 
The parameters determined in this fit are used to determine the peak wavelength in each spectrum from 2014 to 2017 again. 
In this procedure, only the peak wavelength of a Moffat function was set to be free. 
The spectra are shifted to have zero redshift according to the newly determined peak wavelengths, and are then combined.
Here we obtain new combined S0-2 spectrum, and fit it to determine the parameters of the Moffat function. 
The procedure above was repeated iteratively until any of the redshifts for individual years changes no more than 1\,km/s.

\begin{figure}
 \begin{center}
    \includegraphics[width=80mm]{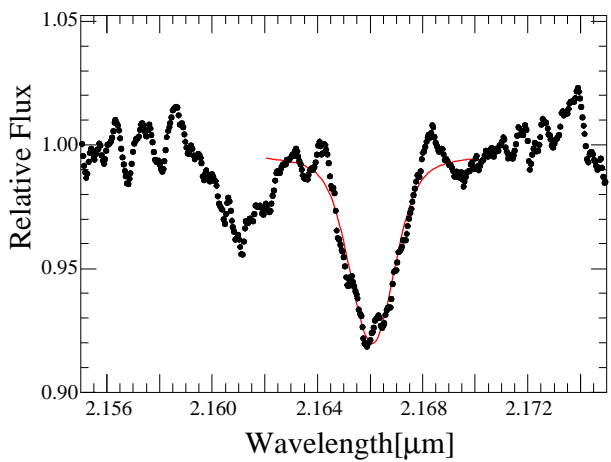}
 \end{center}
\vspace{3cm}
\caption{
Combined S0-2 spectrum ($s = 11$) around the Br-$\gamma$ absorption line, 
obtained with Subaru/IRCS from 2014 to 2017.
We can see a He\,I absorption line at $2.16137\,\mu$m as well as the Br-$\gamma$ $2.16612\,\mu$m line. 
The Br-$\gamma$ line is fitted with a Moffat function (red curve).
}
\label{fig:combinedSpec14-17}
\end{figure}

Figure~\ref{fig:combinedSpec14-17} shows the combined S0-2 spectra around $2.16\,\mu$m, using the Subaru/IRCS data sets from 2014 to 2017.
The total exposure time is 21.8 hours, and the smoothing parameter of $s = 11$.
We can find clearly separated two absorption profiles, He\,I $2.16137\,\mu$m (left) and Br-$\gamma$ $2.16612\,\mu$m (right). 
The Moffat profile used to fit the Br-$\gamma$ $2.16612\,\mu$m line is shown by the red curve.
In the following procedure, this profile will be used to measure the peak wavelength of the Br-$\gamma$ line. 
Only the peak wavelength and scaling factor (corresponds to the continuum level) are set to be free in the following profile fits. 

\subsection{Identification of Br-$\gamma$ feature}
\label{subsec:S02BrgID}

The S0-2 spectra from 2014 May to 2018 Aug obtained with Subaru/IRCS are shown in Fig. \ref{fig:Spec14-17}. 
As shown there, the obtained spectra in 2018 are noisy. 
This is because bad weather conditions, low power output of the LGS system, and frequent satellite closures during the observations in 2018. 
At first glance, it is not clear which feature is the Br-$\gamma$ absorption line of S0-2. 
We therefore carried out an analysis to identify the Br-$\gamma$ absorption before the fitting to determine the redshifts of S0-2.

To identify the feature, we have used the combined spectrum of S0-2 around the Br-$\gamma$ absorption line (Fig. \ref{fig:combinedSpec14-17}). 
By fitting the feature, we have obtained parameters of a Moffat function which fit the feature in the combined spectrum well. 
Using the obtained parameters of the Moffat function for the combined spectrum, we have fitted each spectrum in 2018, by changing the central wavelength of the Moffat function. 
For example, in the case of the 2018 Mar spectra, we have fitted it by changing the central wavelength of the Moffat function from 2.170\,$\mu$m to 2.200\,$\mu$m, and calculate $\chi^2$ values for the fit. 
When we plot $\chi^2$ as a function of the central wavelength, we can find a clear minimum of $\chi^2$ at around $2.194 - 2.195\,\mu$m. 
This suggests that the absorption feature around $2.194 - 2.195\,\mu$m is best matched with the shape of the combined spectrum, compared to other features on the 2018 Mar spectrum.

We have carried out the fitting described above for all the spectra obtained in 2018. 
We have found a clear minimum of $\chi^2$ at 2.194\,$\mu$m, 2.158\,$\mu$m, and 2.153\,$\mu$m for the 2018 Mar, Jul, and Aug spectrum, respectively, 
and thus we have considered the feature at the wavelengths as the Br-$\gamma$ absorption line of S0-2.

For the 2018 May spectrum, we have found two minimums with similar $\chi^2$ values at around 2.178\,$\mu$m and 2.183\,$\mu$m. 
To identify the Br-$\gamma$ feature, we have fitted the redshifts of S0-2 using all the observed ones but that of 2018 May. 
The fitting result suggests that the expected redshift of S0-2 at 2018.382 (2018 May) is $\approx 2630$\,km/s, and the central wavelength of the redshift is $\sim 2.185\,\mu$m. 
Considering the expected redshift,
we have {\it assumed} that the absorption feature at around 2.184\,$\mu$m is the Br-$\gamma$ absorption line of S0-2 at 2018.382. 
Note that without such prediction from other observational results, we cannot distinguish the Br-$\gamma$ line from other features on the 2018 May spectrum. 
Hence the derived uncertainty values for the 2018 May shown below are lower limits of an actual uncertainty in the redshift of S0-2.

\subsection{Redshifts and uncertainties}
\label{subsec:S02ResultsSubaru}

\begin{figure}
 \begin{center}
\includegraphics[width=0.9\textwidth,angle=0]{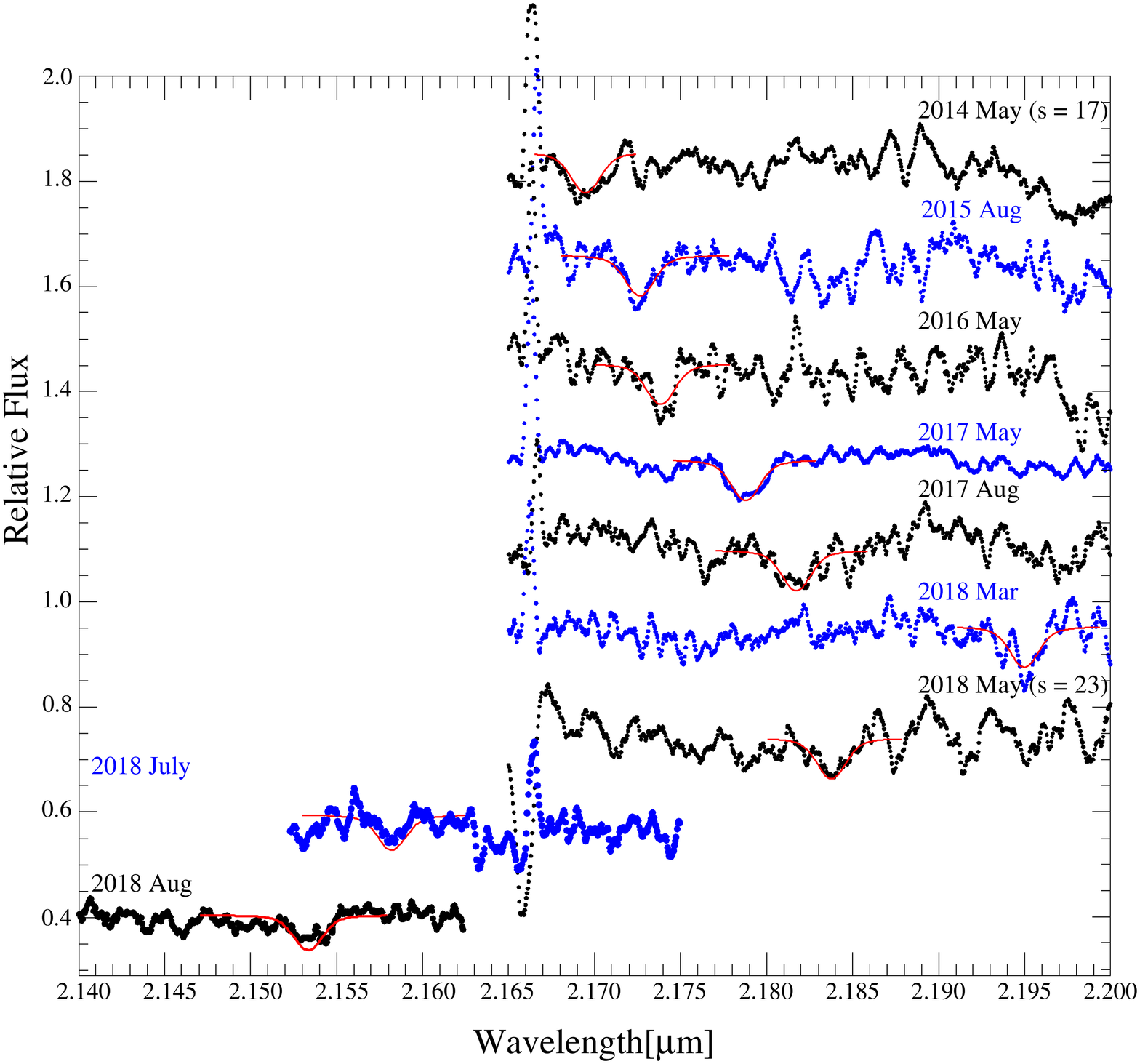}
\caption{
Spectra including Br-$\gamma$ absorption line of S0-2 from 2014 May (top) to 2018 Aug (bottom).
The fitting results are shown by red curves on the spectra.
The smoothing parameters are 17 for 2014, 23 for 2018 May, and 11 for the rest of the spectra.
The LSR correction is not applied.
}
\label{fig:Spec14-17}
\end{center}
\end{figure}

On the S0-2 spectra from 2014 May to 2018 Aug (Fig. \ref{fig:Spec14-17}),
we show the fitting results of the Br-$\gamma$ absorption features using red curves. 
We use the parameters of the Moffat function determined for the combined spectrum (Fig. \ref{fig:combinedSpec14-17}),
but the peak wavelength and the scaling factor (corresponds to the continuum level) are left free in the fits. 
When we fitted the spectra, we divided the 2018 Mar dataset into ``2018 Mar 29 (2018.240)" and ``2018 Mar 30 (2018.243)" datasets. 
The redshifts of S0-2 are determined using the central wavelength of the fitting results, and they are shown in table \ref{tab:subaruUncertainty}.

To determine the uncertainties of the S0-2 redshifts, we have conducted the same procedures described in \citet{ref:nishiyama+2018}. 
To estimate uncertainties, we have carried out Jackknife analysis. 
Before combining observed spectra, we have made $N$ sub-data sets consisting of $N-1$ spectra.
Here $N$ is the number of frames used in data analysis (see Table \ref{Tab:Obs}).
Then we have fitted the Br-$\gamma$ absorption line of the $N$ spectra of the sub-data sets,
and have calculated jackknife uncertainties $\sigma_{\mathrm{JK}}$ using the equation (2) in \citet{ref:nishiyama+2018}.
The obtained jackknife uncertainties are shown in Table \ref{tab:subaruUncertainty}.

Systematic uncertainties $\sigma_{\mathrm{sys}}$ includes the following: 
(1)~uncertainties in spectrum smoothing (typically $1-2\,$km/s); 
(2)~uncertainty in the stability of the long-term wavelength calibration ($\approx 5\,$km/s); 
(3)~uncertainty in the comparison of partly excluded spectra to understand the uncertainty in the telluric correction ($3 - 8\,$km/s).
The spectra used for the fitting (Figs \ref{fig:combinedSpec14-17} and \ref{fig:Spec14-17}) are smoothed one,
because of the faintness of S0-2.
The central wavelengths could have different values when we use different smoothing parameter of the spectra.
Hence we have checked how the central wavelength varies with different smoothing parameters.
The typical uncertainties are estimated to be $1-2\,$km/s.

The systematic uncertainty due to wavelength calibrations, i.e., long-term stability of this spectroscopic monitoring,
is examined using the Br-$\gamma$ ``emission" line.
The interstellar gas around S0-2 is ionized by UV radiation from high-mass stars nearby,
and thus emits Br-$\gamma$ which can be used to estimate the uncertainty of the wavelength calibration.
Assuming the wavelength of the Br-$\gamma$ emission line is stable from 2014 to 2018,
we have fitted the emission line with a Gaussian function and determine the central wavelength for each spectrum.
The standard deviation of the redshifts derived by the central wavelengths are $4.9$\,km/s.

One of the difficulties in data analysis of ground-based near-infrared spectroscopy is removal of telluric absorption features.
In our analysis, we have observed telluric standard stars and used them to remove the telluric lines.
However, the strength and profile of the telluric lines vary with atmospheric conditions and airmass of targets.
Hence we have examined the change of the central wavelengths of the Br-$\gamma$ absorption line by using sub-sets of spectra, part of which is excluded from the original spectra
(for more detail, see \citet{ref:nishiyama+2018}).
In this experiment, we have examined how the central wavelength changes if a part of the Br-$\gamma$ absorption feature
is affected by uncorrected telluric absorption.
The uncertainties derived by the fits of the partly excluded spectra are $3 - 8$\,km/s from 2014 to 2018, 
and these uncertainties are also quadratically added to the final systematic uncertainties of $\sigma_{\mathrm{sys}}$ (Table \ref{tab:subaruUncertainty}).

Note that, as described in section~\ref{subsec:S02BrgID}, 
it is difficult to identify the Br-$\gamma$ absorption feature in the 2018 May spectrum
without a prediction from other data sets.
Hence the uncertainties derived for 2018.382 (Table \ref{tab:subaruUncertainty}) are likely to be
underestimated compared to actual ones.

\begin{table}
\tbl{Redshift and Uncertainties of S0-2 in Subaru/IRCS observations.
}
{
 \begin{tabular}{cccccc}
 \hline
 time\footnotemark[$\ast$] &
 redshift$_{\mathrm{LSR}}$ &
 $\Delta$redshift$_{\mathrm{LSR}}$\footnotemark[$\dag$] &
 $\sigma_{\mathrm{tot}}$ & $\sigma_{\mathrm{JK}}$ &
 $\sigma_{\mathrm{sys}}$
 \\
 yr & km/s & km/s & km/s & km/s & km/s
 \\
 \hline
 $2014.379$ & $\phantom{-0}485.6$ & $+24.6$ & $26.6$ & $25.6$ & $6.6$
 \\
 $2015.635$ & $\phantom{-0}886.6$ & $-15.7$ & $16.5$ & $15.5$ & $5.6$
 \\
 $2016.381$ & $\phantom{-}1096.2$ & $+24.5$ & $16.9$ & $15.2$ & $7.3$
 \\
 $2017.343$ & $\phantom{-}1768.7$ & $+29.9$ & $20.4$ & $20.0$ & $5.9$
 \\
 $2017.348$ & $\phantom{-}1798.8$ & $+29.2$ & $14.4$ & $12.9$ & $6.3$
 \\
 $2017.605$ & $\phantom{-}2133.3$ & $-12.6$ & $27.2$ & $26.1$ & $7.8$
 \\
 $2017.609$ & $\phantom{-}2169.6$ & $ -13.0$ & $36.6$ & $35.7$ & $8.3$
 \\
 $2018.240$ & $\phantom{-}4001.9$ & $+39.6$ & $36.7$ & $35.4$ & $9.5$
 \\
 $2018.243$ & $\phantom{-}4096.6$ & $+39.4$ & $39.6$ & $37.7$ & $12.2$
 \\
 $2018.382$ & $\phantom{-}2466.4$ & $+24.1$ & $67.5$\footnotemark[$\ddag$] & $66.9$\footnotemark[$\ddag$] & $9.1$\footnotemark[$\ddag$]
 \\
 $2018.508$ & $-1102.3$ & $+2.2$ & $53.1$ & $52.5$ & $7.9$
 \\
 $2018.628$ & $-1785.7$ & $-15.0$ & $40.9$ & $39.8$ & $9.3$
 \\
 \hline
 \end{tabular}
}
\label{tab:subaruUncertainty}
\begin{tabnote}
\footnotemark[$\ast$]
Time is counted in the unit of year, setting $1$~yr as $365.25$~days.\\
\footnotemark[$\dag$]
The local standard of rest velocity at the average time of integration.\\
\footnotemark[$\ddag$]
The shown uncertainties for 2018.382 are lower limits.
\end{tabnote}
\end{table}

\onecolumn
%
%
\begin{table}
\caption{Summary of Subaru observations.}
\label{Tab:Obs}
\begin{tabular}{lcccccc}
\hline
 Date &
 Setting\footnotemark[$\mathrm{(a)}$] &
 IT\footnotemark[$\mathrm{(b)}$] &
 $N_{\mathrm{frame}}$\footnotemark[${\mathrm{(c)}}$] &
 $N_{\mathrm{used}}$\footnotemark[${\mathrm{(d)}}$] &
 slit angle\footnotemark[${\mathrm{(e)}}$] &
 AO\footnotemark[${\mathrm{(f)}}$]
\\
 (UTC) & & [sec] & & & [degree]
\\
\hline
 2014 May 19 & $K+$ & 300 & 32 & 30 & 8 & LGS
\\
 2015 Aug 21 & $K+$ & 300 & 24 & 24 & 8 & NGS
\\
 2016 May $18-19$ & $K+$ & 300 & 48 & 44 & 8, 128 & LGS
\\
 2017 May $5-8$ & $K+$ & 300 & 100 & 98 & 8, 117, 127 160, 178 & LGS
\\
 2017 Aug $9-11$ & $K+$ & 300 & 68 & 57 & 8, 127 & NGS/LGS
\\
 2018 Mar $29-30$ & $K+$ & 300 & 39 & 39 & 8, 127 & NGS/LGS
\\
 2018 May $20$ & $K+$ & 300 & 34 & 32 & 68 & NGS
\\
 2018 Jul $4-6$ & $K-$ & 300 & 48 & 42 & 70, 117, 160 & NGS/LGS
\\
 2018 Aug $18$ & $K-$ & 300 & 24 & 24 & 8, 70 & NGS
\\
\hline
\end{tabular}
\begin{tabnote}
\footnotemark[$\mathrm{(a)}$]
 IRCS Echelle setting.
\\
\footnotemark[$\mathrm{(b)}$]
 Integration time for each exposure.
\\
\footnotemark[$\mathrm{(c)}$]
 The number of frames taken in the night(s)
\\
\footnotemark[$\mathrm{(d)}$]
 The number of frames used in data analysis.
\\
\footnotemark[$\mathrm{(e)}$]
 The angular offset measured from north to east, counterclockwise.
\\
\footnotemark[$\mathrm{(f)}$]
 The guide star of the AO system. The ``LGS" mode uses the laser guide star system, and the ``NGS" mode uses only a natural guide star.
\end{tabnote}
\end{table}

\twocolumn

\section{Time evolution of $\Delta z\ls{1PN.0PM}(t)$ fitted with observational data}
\label{sec:fitting}

As derived in section~\ref{sec:theory}, the difference between the GR and Newtonian redshifts under our presupposition $\Delta z\ls{1PN.0PM}(t)$ shows the ``double-peak-appearance'' in its time evolution. 
In this section, we examine whether the double-peak-appearance is found or not in the observational data including our 2018 data.

Note that the observational data used in our analysis include not only our own data but also all public data released by the other groups by 2017, while VLT group~\citep{ref:gravity+2018} did not use the astrometric data of Keck group~\citep{ref:do+2019}, and Keck group did not use the astrometric data of VLT group. 
Further, the new 2018 data in \citet{ref:gravity+2018} and \citet{ref:do+2019} are not used in our analysis, because those data are not available for us when this paper was written.

\subsection{Parameters determined by our fitting}
\label{sec:fitting.parameter}

The parameters determined by the $\chi^2$-fitting in the following discussions are not only the eleven parameters listed in section~\ref{sec:theory.evolution.step1} but also the parameters corresponding to the origin of the astrometric data of Keck and VLT groups. 
Their astrometric origins are set at the position of an infra-red-flare near \sgra at a certain time. 
The flare position at a certain time may be moving relative to \sgra (and to our astrometric center ``C'' introduced in appendix~\ref{app:coordinate}). 
Therefore, the vector $\vec{A}\ls{sky}(t)$ defined in equation~\eqref{eq:coordinate.Asky} of appendix~\ref{app:coordinate} does not vanish for either Keck or VLT astrometric data. 
Then, we introduce the following eight parameters corresponding to $\vec{A}\ls{sky}(t)$:
\begin{list}{}{}
\item[$\Delta{\rm RA}\ls{K.apo}$:]
the R.A. of the astrometric origin $\vec{A}\ls{apo}$ at the apocenter time $t\ls{S.apo}$ for the Keck data.
\item[$\Delta{\rm DEC}\ls{K.apo}$:]
the Dec. of the astrometric origin $\vec{A}\ls{apo}$ at the apocenter time $t\ls{S.apo}$ for the Keck data.
\item[$V\ls{K.ra}$:]
the R.A.-component of the velocity of astrometric origin $\vec{V}\ls{astro}$ for the Keck data.
\item[$V\ls{K.dec}$:]
the Dec.-component of the velocity of astrometric origin $\vec{V}\ls{astro}$ for the Keck data.
\item[$\Delta{\rm RA}\ls{V.apo}$:]
the R.A. of the astrometric origin $\vec{A}\ls{apo}$ at the apocenter time $t\ls{S.apo}$ for the VLT data.
\item[$\Delta{\rm DEC}\ls{V.apo}$:]
the Dec. of the astrometric origin $\vec{A}\ls{apo}$ at the apocenter time $t\ls{S.apo}$ for the VLT data.
\item[$V\ls{V.ra}$:]
the R.A.-component of the velocity of astrometric origin $\vec{V}\ls{astro}$ for the VLT data.
\item[$V\ls{V.dec}$:]
the Dec.-component of the velocity of astrometric origin $\vec{V}\ls{astro}$ for the VLT data.
\end{list}
In total, we determine the nineteen parameters by the $\chi^2$-fitting of the S0-2 motion with real astrometric and spectroscopic observational data.

\subsection{Results of fitting}
\label{sec:fitting.result}

\begin{longtable}{c||c|c|c|c}
\caption{
Results of our $\chi^2$-fitting. 
GR-best-fit is the result of fitting the real observational data with the S0-2 motion at the 1PN+0PM approximation of GR. 
NG-best-fit is the result of fitting the real observational data with the S0-2 motion in the Newtonian gravity. 
NG-art-best-fit is the result of fitting the artificial accurate data with the S0-2 motion in the Newtonian gravity, where the artificial data are created from the GR-best-fit. 
The error in $\chi^2$-fitting is given by definition~\eqref{eq:theory.formalerror}.
}
\label{tab:fitting.result}
\endfirsthead 
\hline
 $\rcsmin$ and parameters
 &
 $\rcsmin$ &
 $M\ls{SgrA}$ &
 $R\ls{GC}$ &
 $V\ls{O.ra}$
\\
 determined by $\chi^2$-fitting
 &
 [no dim.] &
 [$10^6 M_\odot$] &
 [kpc] &
 [mas/yr]
\\
\hline
 GR-best-fit
 &
 1.1903 &
 $4.232\pm 0.066$ &
 $8.098\pm 0.066$ &
 $-0.162\pm 28.782$
\\
\hline
 NG-best-fit
 &
 1.2134 &
 $4.274\pm 0.067$ &
 $8.114\pm 0.067$ &
 $-0.168\pm 29.056$
\\
\hline
 NG-art-best-fit
 &
 0.0754 &
 $4.352\pm 0.020$ &
 $8.207\pm 0.018$ &
 $-0.128\pm 0.002$
\\
\hline\hline
Parameters
 &
 $V\ls{O.dec}$ &
 $V\ls{O.Z}$ &
 $I\ls{S}$ &
 $\Omega\ls{S}$
\\
 &
 [mas/yr] &
 [km/s] &
 [deg] &
 [deg]
\\
\hline
 GR-best-fit
 &
 $0.174\pm 28.787$ &
 $-8.345\pm 3.213$ &
 $134.239\pm 0.217$ &
 $227.766\pm 0.242$
\\
\hline
 NG-best-fit
 &
 $0.166\pm 29.061$ &
 $5.261\pm 3.196$ &
 $134.063\pm 0.214$ &
 $227.518\pm 0.245$
\\
\hline
 NG-art-best-fit
 &
 $0.191\pm 0.001$ &
 $9.307\pm 1.070$ &
 $134.306\pm 0.056$ &
 $226.974\pm 0.066$
\\
\hline\hline
Parameters
 &
 $\omega\ls{S}$ &
 $e\ls{S}$ &
 $T\ls{S}$ &
 $t\ls{S.apo}$
\\
 &
 [deg] &
 [no dim.] &
 [yr] &
 [AD]
\\
\hline
 GR-best-fit
 &
 $66.204\pm 0.333$ &
 $0.8903\pm 0.0007$ &
 $16.0504\pm 0.0023$ &
 $2010.3383\pm 0.0015$
\\
\hline
 NG-best-fit
 &
 $66.049\pm 0.340$ &
 $0.8911\pm 0.0007$ &
 $16.0468\pm 0.0022$ &
 $2010.3432\pm 0.0014$
 \\
\hline
 NG-art-best-fit
 &
 $65.521\pm 0.062$ &
 $0.8899\pm 0.0002$ &
 $16.0489\pm 0.0003$ &
 $2010.3410\pm 0.0003$
\\
\hline\hline
Parameters
 &
 $\Delta{\rm RA}\ls{K.apo}$ &
 $\Delta{\rm DEC}\ls{K.apo}$ &
 $V\ls{K.ra}$ &
 $V\ls{K.dec}$
\\
 &
 [mas] &
 [mas] &
 [mas/yr] &
 [mas/yr]
\\
\hline
 GR-best-fit
 &
 $0.576\pm 0.611$ &
 $-1.796\pm 0.611$ &
 $0.262\pm 28.790$ &
 $-0.708\pm 28.790$
\\
\hline
 NG-best-fit
 &
 $0.689\pm 0.620$ &
 $-1.725\pm 0.620$ &
 $0.309\pm 29.063$ &
 $-0.692\pm 29.063$
 \\
\hline\hline
Parameters
 &
 $\Delta{\rm RA}\ls{V.apo}$ &
 $\Delta{\rm DEC}\ls{V.apo}$ &
 $V\ls{V.ra}$ &
 $V\ls{V.dec}$
\\
 &
 [mas] &
 [mas] &
 [mas/yr] &
 [mas/yr]
\\
\hline
 GR-best-fit
 &
 $-1.061\pm 0.611$ &
 $2.152\pm 0.611$ &
 $0.154\pm28.790$ &
 $-0.220\pm 28.790$
\\
\hline
 NG-best-fit
 &
 $-0.964\pm 0.620$ &
 $2.223\pm 0.620$ &
 $0.200\pm 29.063$ &
 $-0.199\pm 29.063$
 \\
\hline
\end{longtable}

As the first step, we perform three $\chi^2$-fittings in order to obtain three sets of parameter values:
\begin{list}{}{}
\item[Fitting 1 (GR-best-fit):]
We perform the $\chi^2$-fitting of the real observational data with the S0-2 motion at the 1PN+0PM approximation. 
Then we obtain the \emph{GR best-fitting values of the nineteen parameters}, which are shown in table~\ref{tab:fitting.result} as ``GR-best-fit''. 
With these parameter values, the redshift at the 1PN+0PM approximation, $cz\ls{1PN.0PM}(t)$, is calculated using equation~\eqref{eq:theory.z1PN0PM}. 
\item[Fitting 2 (NG-best-fit):]
We perform the $\chi^2$-fitting of the real observational data with the S0-2 motion in the Newtonian gravity. 
Then we obtain the \emph{NG best-fitting values of the nineteen parameters}, which are shown in table~\ref{tab:fitting.result} as ``NG-best-fit''. 
With these parameter values, the redshift in the Newtonian gravity, $cz\ls{NG}\us{(real)}(t)$, is calculated using equation~\eqref{eq:theory.zNG}. 
Here the upper suffix ``(real)'' denotes that this Newtonian redshift is obtained from the real observational data. 
\item[Fitting 3 (NG-art-best-fit):]
We create the ideally accurate, artificial data set using the GR-best-fit values of the eleven parameters listed in section~\ref{sec:theory.evolution.step1}, where we set $N=20$ and $L=64$~yr, which are the parameters introduced in section~\ref{sec:theory.evolution.step2}.\footnote{
The values of $N=20$ and $L=64$~yr are one example. 
The other cases satisfying the conditions given in section~\ref{sec:theory.evolution.step2} result in the same conclusion with this section.
} 
Then, we perform the $\chi^2$-fitting of this artificial data set with the S0-2 motion in the Newtonian gravity, and we obtain the \emph{NG artificial best-fitting values of the eleven parameters}, which are shown in table~\ref{tab:fitting.result} as ``NG-art-best-fit''. 
With these parameter values, the redshift in the Newtonian gravity, $cz\ls{NG}\us{(art)}(t)$, is calculated using equation~\eqref{eq:theory.zNG}. 
Here the upper suffix ``(art)'' denotes that this Newtonian redshift is obtained from the artificial data set. 
\end{list}

\begin{figure}
 \begin{center}
 \includegraphics[width=80mm]{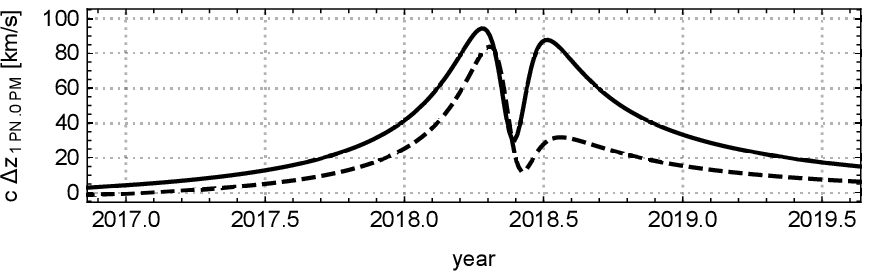} 
\\
 \includegraphics[width=80mm]{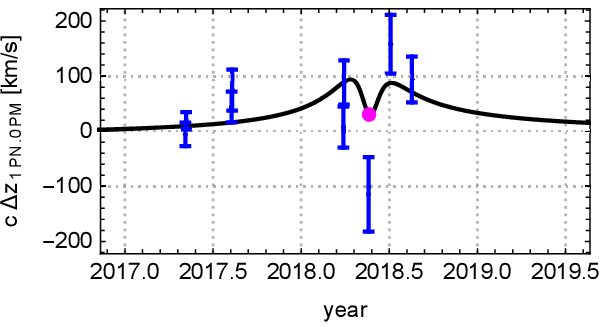} 
\\
 \includegraphics[width=80mm]{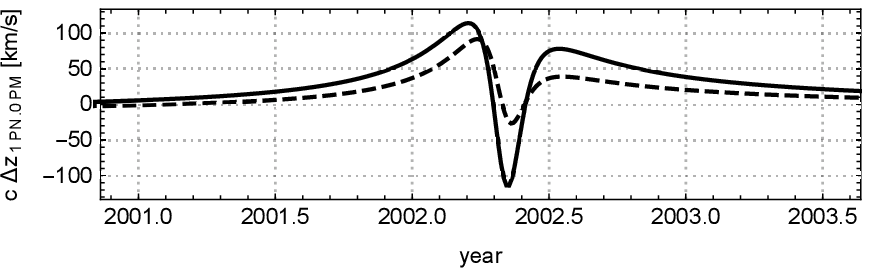} 
\\
 \includegraphics[width=80mm]{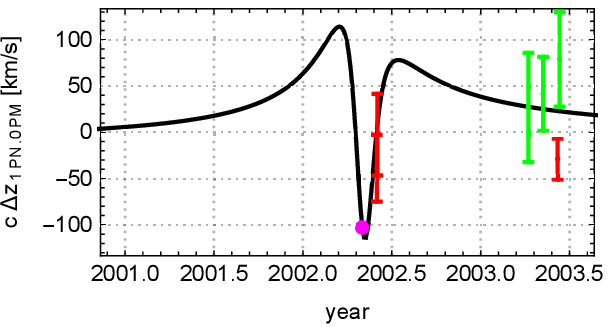} 
\\
 \includegraphics[width=80mm]{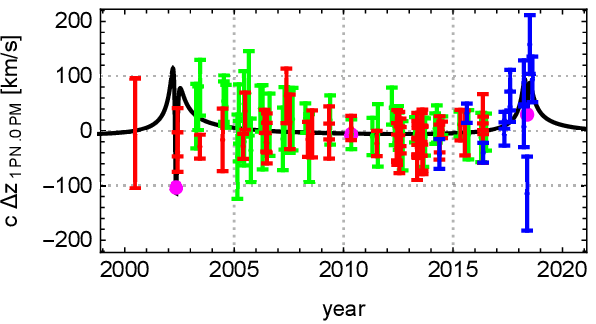}
 \end{center}
\caption{
GR evidence described by our fitting results. 
In all panels, the solid curve is the real GR evidence, $c\Delta z\ls{1PN.0PM}\us{(real)}(t)$, obtained from the real observational data, and the dashed curve is the theoretically expected GR evidence, $c\Delta z\ls{1PN.0PM}\us{(art)}(t)$. 
The blue data are of Subaru observations, red data of Keck, and green data of VLT, where all data are subtracted by the Newtonian best-fitting redshift $cz\ls{NG}\us{(real)}(t)$, see equation~\eqref{eq:theory.dzobs}. 
Magenta dots denote the pericenter and apocenter passages. 
The 1st and 2nd panels focus around the recent pericenter passage. 
The 3rd and 4th panels focus around the previous pericenter passage. 
The 5th panel shows the temporal range including all real observational data. 
}
\label{fig:fitting.GReffect}
\end{figure}

Next, we calculate the following two types of the measure of GR evidence~\eqref{eq:theory.dz1PN0PM} under our presupposition on the parameter values:
\eqab
\label{eq:fitting.dz.observe}
 c\Delta z\ls{1PN.0PM}\us{(observe)}(t)
 &\defeq&
 cz\ls{1PN.0PM}(t) - cz\ls{NG}\us{(real)}(t)
 \\
\label{eq:fitting.dz.expect}
 c\Delta z\ls{1PN.0PM}\us{(expect)}(t)
 &\defeq&
 cz\ls{1PN.0PM}(t) - cz\ls{NG}\us{(art)}(t)
 \,.
\eqae
The former, $c\Delta z\ls{1PN.0PM}\us{(observe)}(t)$, is the observed GR evidence estimated from only real observational data. 
The latter, $c\Delta z\ls{1PN.0PM}\us{(expect)}(t)$, is the theoretically expected form of the GR evidence, under the assumption that the GR-best-fit parameter values represent the true S0-2 motion. 
As discussed in section~\ref{sec:theory.deviation}, if the time evolution of $c\Delta z\ls{1PN.0PM}\us{(observe)}(t)$ matches well with that of $c\Delta z\ls{1PN.0PM}\us{(expect)}(t)$, then it is concluded that the real observational data are described well by GR.

Figure~\ref{fig:fitting.GReffect} shows the GR evidence represented by our fitting results listed in table~\ref{tab:fitting.result}. 
The first and second panels focus around the recent pericenter passage, where the solid curve is the time evolution of the observed GR evidence $c\Delta z\ls{1PN.0PM}\us{(observe)}(t)$ and the dashed curve is the theoretically expected time evolution of the GR evidence $c\Delta z\ls{1PN.0PM}\us{(expect)}(t)$. 
The third and fourth panels focus around the previous pericenter passage. 
The fifth panel shows the whole temporal range covering all real observational data, in which the Subaru data are denoted by blue, the Keck data by red, and the VLT data by green. 
Those data points are $c\Delta z\ls{obs}$, defined in equation~\eqref{eq:theory.dzobs}.

The double-peak-appearance around the recent and previous pericenter passages are recognized in the observed GR evidence (solid curve in figure~\ref{fig:fitting.GReffect}). 
In order for a quantitative assessment of the detection of the GR evidence under our presupposition on the parameter values, we need a quantity that can measure the discrepancy/similarity between the solid curve and the dashed curve in figure~\ref{fig:fitting.GReffect}. 
Such a quantity is defined in section~\ref{sec:fitting.measure}.

\begin{figure}
 \begin{center}
 \includegraphics[width=80mm]{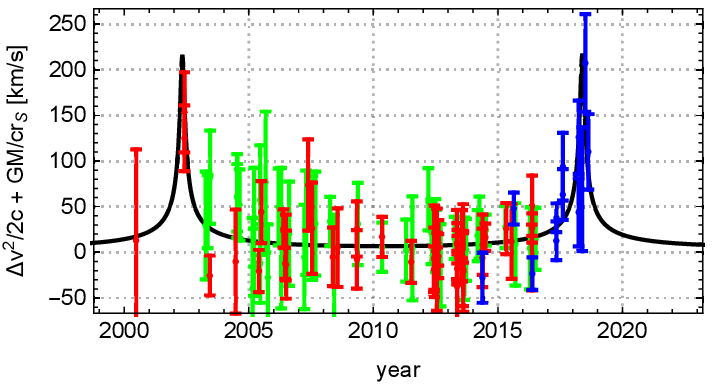}
\\
 \includegraphics[width=80mm]{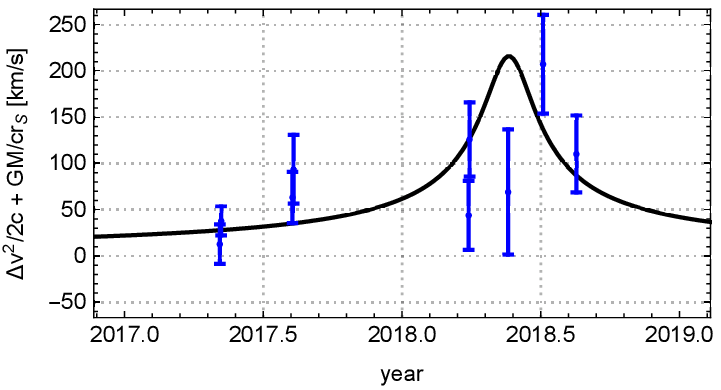}
 \end{center}
\caption{
Previously-used GR evidence given by equation~\eqref{eq:theory.SRGRterm} with GR-best-fit parameters. 
The top panel shows the temporal range for all observational data. 
The bottom panel focuses around the recent pericenter passage. 
The blue data are of Subaru observations, red data of Keck, and green data of VLT, where all data are subtracted by the LSV part of GR redshift (1st and 2nd terms of equation~\eqref{eq:theory.z1PN0PM}). 
}
\label{fig:fitting.Formal}
\end{figure}

\begin{figure}
 \begin{center}
 \includegraphics[width=80mm]{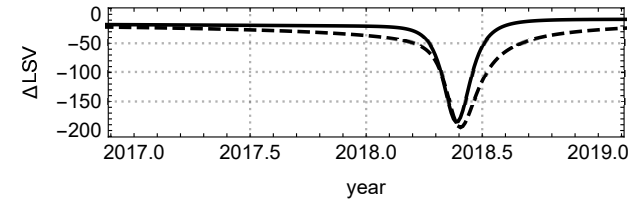} 
 \end{center}
\caption{Difference of LSV between 1PN+0PM (GR-best-fit) and Newtonian cases. 
For the solid curve, the Newtonian case is given by the NG-best-fit parameters. 
For the dashed curve, the Newtonian case is the NG-art-best-fit parameters, corresponding to the bottom panel of figure~\ref{fig:theory.GReffect.component}. 
}
\label{fig:fitting.RVdiff}
\end{figure}

Here we summarize some points found in the figures~\ref{fig:fitting.Formal} and~\ref{fig:fitting.RVdiff}. 
Figure~\ref{fig:fitting.Formal} shows the ``previously-used'' GR evidence (the special relativistic and gravitational Doppler components in $\Delta z\ls{1PN.0PM}$) given by the formula~\eqref{eq:theory.SRGRterm}, which corresponds to the case of $f=1$ of $\Delta z\ls{GR}\us{(prev)}$ in equation~\eqref{eq:theory.dzprev}. 
We find that our largest magnitude of the previously-used GR evidence $\approx 200$ km/s appears around the pericenter passage (2018.3850~yr in 1PN+0PM motion of S0-2). 
This is consistent with the results of the other groups.

Note that the largest magnitude of $\Delta z\ls{1PN.0PM} \approx 100$ km/s under our presupposition (see figure~\ref{fig:fitting.GReffect}) is about a half of that of the previously-used GR evidence $\approx 200$ km/s. 
Even when the previously-used GR evidence would be detected with a given observational data set, the significance for the detection of $\Delta z\ls{1PN.0PM}$ would be smaller under our presupposition.

Figure~\ref{fig:fitting.RVdiff} shows the difference of the LSV between the 1PN+0PM motion of S0-2 (GR-best-fit parameters) and the Newtonian motion of S0-2 (NG-best-fit or NG-art-best-fit), given by equation~\eqref{eq:theory.LSVterm}. 
For the solid curve, the Newtonian motion is given by the NG-best-fit parameters. 
For the dashed curve, the Newtonian motion is the NG-art-best-fit parameters (corresponding to the bottom panel of figure~\ref{fig:theory.GReffect.component}). 
The negativity of both solid and dashed curves denotes that the Newtonian LSV is faster than the GR LSV in both NG-best-fit and NG-art-best-fit. 
Further, from figures~\ref{fig:fitting.Formal} and~\ref{fig:fitting.RVdiff}, it is recognized that the summation of the previously-used GR evidence and the LSV difference results in the double-peak-appearance of our GR evidence $\Delta z\ls{1PN.0PM}(t)$ shown in figure~\ref{fig:fitting.GReffect}. 
This is consistent with the simulation (figure~\ref{fig:theory.GReffect.component}) performed in section~\ref{sec:theory.evolution.step4}.

\subsection{A quantity to measure the discrepancy between the GR and the Newtonian gravity}
\label{sec:fitting.measure}

As discussed in the second paragraph of section~\ref{sec:theory.evolution.step3}, the $\chi^2$-assessment within the present observational precision is not useful for detecting the discrepancy between the GR and the Newtonian gravity. 
Further, we do not introduce any auxiliary parameter (see section~\ref{sec:theory.deviation}) under our presupposition on the parameter values. 
Then, instead of the $\rcs$ whose definition is based mainly on the statistical mathematics, we define the following quantity, $\delta z$, that is based mainly on the double-peak-appearance of the GR evidence under our presupposition:
\eqb
 \delta z(t_0,\delta t)
 \defeq
 \displaystyle
 \frac{
 \displaystyle
 \int_{t_0}^{t_0+\delta t}
 \diff{t}
 \left| \Delta z\ls{1PN.0PM}\us{(observe)}(t) - \Delta z\ls{1PN.0PM}\us{(expect)}(t)
 \right|
 }
 {
 \displaystyle
 \int_{t_0}^{t_0+\delta t}
 \diff{t}
 \left| \Delta z\ls{1PN.0PM}\us{(expect)}(t)
 \right|
 }
\label{eq:fitting.measure}
 \,,
\eqe
where $t_0$ and $\delta t$ have the dimension of time. 
The interpretation of this definition~\eqref{eq:fitting.measure} is as follows:
\begin{itemize}
\item
The denominator of $\delta z(t_0,\delta t)$ represents an absolute amount of the theoretically expected GR evidence $\Delta z\ls{1PN.0PM}\us{(expect)}(t)$ for duration $\delta t$ from $t_0$. 
For example in the first panel of figure~\ref{fig:fitting.GReffect}, the denominator corresponds to the area between the horizontal axis and the curve of $c \Delta z\ls{1PN.0PM}\us{(expect)}(t)$ (dashed curve). 
\item
The numerator of $\delta z(t_0,\delta t)$ represents an absolute amount of the difference between the observed GR evidence $\Delta z\ls{1PN.0PM}\us{(observe)}$ and the theoretically expected one $\Delta z\ls{1PN.0PM}\us{(expect)}$ for duration $\delta t$ from $t_0$. 
For example in the first panel of figure~\ref{fig:fitting.GReffect}, the numerator corresponds to the area between the curve of $c \Delta z\ls{1PN.0PM}\us{(observe)}(t)$ (solid curve) and the curve of $c \Delta z\ls{1PN.0PM}\us{(expect)}(t)$ (dashed curve). 
Note the mathematical fact that, if the real observational data expresses exactly the theoretically expected time evolution of the GR evidence under our presupposition, then the numerator of $\delta z$ must be zero. 
\item
The quantity $\delta z(t_0,\delta t)$ defined in equation~\eqref{eq:fitting.measure} is the ratio of the ``difference between $\Delta z\ls{1PN.0PM}\us{(observe)}$ and $\Delta z\ls{1PN.0PM}\us{(expect)}$'' to the ``amount of $\Delta z\ls{1PN.0PM}\us{(expect)}$'', for duration $\delta t$ from $t_0$. 
In other words, this $\delta z$ expresses to what extent the GR evidence in the real observational data matches well with the theoretically expected GR evidence under our presupposition on the parameter values. 
\end{itemize}
We propose this $\delta z(t_0,\delta t)$ as a measure of the discrepancy/similarity between the GR and the Newtonian gravity under our presupposition on the parameter values.

To calculate the value of $\delta z(t_0,\delta t)$, we need to determine not only the values of $t_0$ and $\delta t$ but also the three sets of parameters, GR-best-fit, NG-best-fit and NG-art-best-fit. 
Using the three sets of best-fitting parameters listed in table~\ref{tab:fitting.result}, some values of $\delta z(t_0,\delta t)$ for some combinations of $(t_0,\delta t)$ are listed in table~\ref{tab:fitting.error}. 
This table implies that the present real observational data include about $60\%$ error in measuring the double-peak-appearance of the GR evidence. 
In order to reduce this error and to confirm the detection of the double-peak-appearance, we need additional data sets.

\begin{table}
\tbl{The error, $\delta z(t_0,\delta t)$, that estimates a discrepancy between the observed GR evidence $\Delta z\ls{1PN.0PM}\us{(observe)}$ and the theoretically expected GR evidence $\Delta z\ls{1PN.0PM}\us{(expect)}$ for the real observational data set. 
}
{
 \begin{tabular}{ccc}
 \hline
 $\delta z(t_0,\delta t)$ & counted from $t_0$ [yr] & with duration $\delta t$ [yr]
 \\
 \hline
 $0.651$ & $2000.475$\footnotemark[$\ast$] & $18.135$\footnotemark[$\dag$]
 \\
 $0.603$ & $2018.628-T\ls{S}$\footnotemark[$\ddag$] & $T\ls{S}$
 \\
 $0.623$ & $t\ls{S.apo}$ & $T\ls{S}$
 \\
 $0.636$ & $2000.475$ & $T\ls{S}$
 \\
 $0.620$ & $t\ls{S.apo}-T\ls{S}$\footnotemark[$\sharp$] & $T\ls{S}$
 \\
 $0.584$ & $t\ls{S.apo}-\frac{T\ls{S}}{2}$\footnotemark[$\flat$] & $T\ls{S}$
 \\
 \hline
 \end{tabular}
}
\label{tab:fitting.error}
\begin{tabnote}
\footnotemark[$*$]
The first spectroscopic data was observed at $2000.475$ by Keck, and the latest data was at 2018.628 by Subaru shown in table~\ref{tab:subaruUncertainty}.
\\
\footnotemark[$\dag$]
Temporal range of the real spectroscopic data is $\delta t = 2018.628-2000.475 = 18.135$ yr.
\\
\footnotemark[$\ddag$]
$T\ls{S}$ and $t\ls{S.apo}$ are in the GR-best-fit parameters in table~\ref{tab:fitting.result}, that correspond to, respectively, the recent apocenter time in 2010 and the period of the S0-2 motion. 
\\
\footnotemark[$\sharp$]
$t\ls{S.apo}-T\ls{S}$ is about the time at previous apocenter in 1994.
\\
\footnotemark[$\flat$]
$t\ls{S.apo}-T\ls{S}/2$ is about the time at previous pericenter in 2002.
\end{tabnote}
\end{table}

\section{Summary and discussion}
\label{sec:sd}

Under the presupposition on the parameter values given in section~\ref{sec:intro}, we have proposed a theoretical discussion on the GR evidence that appears in the spectroscopic data of the S0-2 motion. 
The GR evidence under our presupposition, $\Delta z\ls{1PN.0PM}(t)$ defined in equation~\eqref{eq:theory.dz1PN0PM}, is the difference between the GR and Newtonian redshifts of photons coming from S0-2. 
In section~\ref{sec:theory}, under our presupposition, we have revealed that the theoretically expected time evolution of $\Delta z\ls{1PN.0PM}(t)$ shows two peaks, before and after the pericenter passage of S0-2. 
This ``double-peak-appearance'' is a significant feature which expresses the discrepancy between GR and Newtonian gravity under our presupposition on the parameter values. 
(The double peaks reduce to a single peak under the treatment of the parameter values by the other groups, as summarized in section~\ref{sec:theory.deviation}.) 
It has also been found that the $\chi^2$-assessment under the present averaged observational uncertainties is not useful to confirm the detection of the double-peak-appearance.

In section~\ref{sec:data}, our observations with the Subaru telescope by 2018 have been summarized. 
Due to unfortunate bad weather conditions at Hawaii island in 2018, our data in 2018 have lower SN ratio than the previous years. 
In section~\ref{sec:fitting}, it has been shown that the double-peak-appearance can be recognized in the present observational data (figure~\ref{fig:fitting.GReffect}). 
However, as shown in figure~\ref{fig:fitting.GReffect}, the uncertainties of our 2018 data are so large that we cannot exclude the Newtonian case ($z\ls{obs} \equiv z\ls{NG}$). 
Further, according to the quantity $\delta z$ which measures the discrepancy between the GR and the Newtonian gravity under our presupposition on the parameter values, the error in measuring the double-peak-appearance in the present data set is about $60\%$. 
In order to reduce this error and to confirm the detection of the double-peak-appearance, we need additional data sets.

Finally, let us discuss one method for the test of GR or the so-called modified theories of gravity, under our presupposition on the parameter values. 
Note again that the quantity $\delta z$ estimates the discrepancy between the GR and the Newtonian gravity under our presupposition. 
Therefore, if we replace the GR with a modified theory of gravity in the definition of $\delta z$, then the modified $\delta z$ can be interpreted as a measure of the discrepancy between the modified theory of gravity and the Newtonian gravity. 
Hence, if the value of $\delta z$ of the GR is lower than the value of $\delta z$ of the other theories of gravity, then it is reasonable to conclude that the GR is more promising than the other theories of gravity.

\begin{ack}
We would like to express our gratitude to the staffs of the Subaru telescope, for their continuous supports for our observations. 
We thank Rainer Sch\"{o}del for his supports in our data analysis, and Aurelien Hees for his useful discussions on the theory for detecting the GR evidence. 
H.~S. was supported by JSPS KAKENHI, Grant-in-Aid for Challenging Exploratory Research 26610050, and Grant-in-Aid for Scientific Research (B) 19H01900. 
S.~N. was supported by JSPS KAKENHI, Grant-in-Aid for Young Scientists (A) 25707012, Grant-in-Aid for Challenging Exploratory Research 15K13463 and 18K18760, and Grant-in-Aid for Scientific Research (A) 19H00695. 
T.~O. was supported by JSPS KAKENHI, Grant-in-Aid for JSPS fellows JP17J00547. 
Y.~T. was supported by JSPS KAKENHI, Grant-in-Aid for Young Scientists (B) 26800150. 
M.~T. was supported by DAIKO FOUNDATION, and JSPS KAKENHI, Grant-in-Aid for Scientific Research (C) 17K05439.
\end{ack}

\appendix 
\section{Set-up of our coordinate system}
\label{app:coordinate}

\begin{figure}
 \rightline{
 \includegraphics[width=70mm]{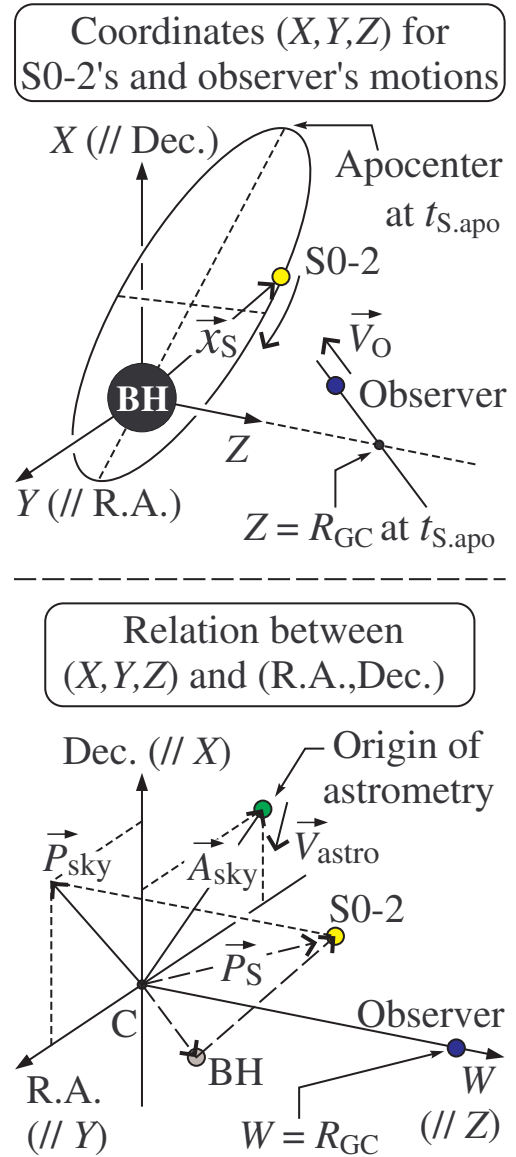}
 }
\caption{
Top panel denotes the coordinate system $(X,Y,Z)$ for calculating the motions of S0-2 and observer. 
The origin of $(X,Y,Z)$ is fixed at \sgra, and the origin of time is set at the apocenter time of S0-2, $t\ls{S.apo}$. 
Bottom panel shows the relation between $(X,Y,Z)$ and the astrometric coordinate system, right ascension (R.A.) and declination (Dec.). 
The center ``C'' of R.A. and Dec. is at distance $R\ls{GC}$ in the direction parallel to $Z$-axis, and the origin of astrometric observations is, in general, moving relative to C.
}
\label{fig:theory.coordinate}
\end{figure}

In calculating the motion of S0-2, we fix the coordinate system, $(t,r,\theta,\varphi)$, centered at \sgra. 
For the Newtonian case, the spatial coordinates $(r,\theta,\varphi)$ are the usual spherical polar coordinates. 
For the GR case, the spacetime coordinates $(t,r,\theta,\varphi)$ are the Boyer-Lindquist coordinates in Kerr spacetime, although the spacetime metric tensor will be approximated to be the 1st order post-Newtonian (1PN) form in section~\ref{sec:theory.approximation}. 
In both GR and Newtonian cases, the origin of time is set at the apocenter passage of S0-2 in 2010. 
Further, due to the huge distance between Sun and \sgra $\approx 8$~kpc, we assume that the relative velocity of the observer measured from \sgra, $\vec{V}\ls{O}$, is constant. 
This coordinate system set-up is schematically shown in the top panel of figure~\ref{fig:theory.coordinate}. 
In this figure, the direction of BH spin is ignored because we consider the 1PN form (the effect of BH spin is ignored) in the main text of this paper, and the coordinates $(X,Y,Z)$ are given by
\eqb
\label{eq:coordinate.XYW}
  X = r \sin\theta \cos\varphi \,,\,
  Y = r \sin\theta \sin\varphi \,,\,
  Z = r \cos\theta \,,
\eqe
where the $Z$-axis points to us (observer) from \sgra (BH) at the apocenter time $t\ls{S.apo}$, and the directions of $X$ and $Y$ axes are, respectively, parallel to the directions of declination (Dec.) and right ascension (R.A.). 
The line-of-sight direction, which points to S0-2 from the observer, changes due to the motions of S0-2 and the observer.

In combining our numerical calculation with the astrometric observational data of S0-2, we use the relation between $(X,Y,Z)$ and the astrometric coordinates (R.A. and Dec.) as shown in the bottom panel of figure~\ref{fig:theory.coordinate}. 
We put the center ``C'' of R.A. and Dec. axes at distance $R\ls{GC}$ from the observer in the direction parallel to $Z$-axis. 
The $W$-axis points to the observer from C, and its direction is parallel to $Z$-axis. 
The spatial position of S0-2 measured from C at a time $t$, $\vec{P}\ls{S}(t)$, is give by
\eqb
\label{eq:coordinate.PS}
 \vec{P}\ls{S}(t) = \vec{x}\ls{S}(t) - (t-t\ls{S.apo}) \vec{V}\ls{O} \,,
\eqe
where $\vec{x}\ls{S}(t)$ is the spatial position of S0-2 measured from \sgra at time $t$, and the second term is the spatial position of \sgra measured from C at time $t$. 
Our definition of R.A. and Dec. of S0-2 is given by the projection of $\vec{P}\ls{S}(t)$ onto the sky-plane, $\vec{P}\ls{sky}(t)$.

In real astrometric observations, the origin of astrometry is not necessarily the same as our center C, and given by the following vector on the sky-plane,
\eqb
\label{eq:coordinate.Asky}
 \vec{A}\ls{sky}(t) = \vec{A}\ls{apo} + (t-t\ls{S.apo}) \vec{V}\ls{astro} \,,
\eqe
where $\vec{A}\ls{apo}$ is the position of the astrometric origin at the apocenter time $t\ls{S.apo}$, and $\vec{V}\ls{astro}$ is the relative velocity of the astrometric origin measured from C. 
Here we assume $\vec{V}\ls{astro}$ is constant. 
The real astrometric observational data are compared with the numerical values of R.A. and Dec. of S0-2 given by $\vec{P}\ls{sky}(t) - \vec{A}\ls{sky}(t)$,
\eqab
\label{eq:coordinate.RA}
 \mbox{R.A. of S0-2} &\,:\,&
 \arctan\Bigl(\frac{P\ls{sky.R}-A\ls{sky.R}}{R\ls{GC}-P\ls{S.W}}\Bigr)
\\
\label{eq:coordinate.Dec}
 \mbox{Dec. of S0-2} &\,:\,& 
 \arctan\Bigl(\frac{P\ls{sky.D}-A\ls{sky.D}}{R\ls{GC}-P\ls{S.W}}\Bigr)
 \,,
\eqae
where the suffices ``R'', ``D'' and ``W'' denote, respectively, the components of vector in R.A., Dec. and $W$ directions.

Finally we make two comments. 
First, because the center C at the apocenter time $t\ls{S.apo}$ is just at \sgra (the origin of coordinates $(X,Y,Z)$\,), the first term $\vec{A}\ls{apo}$ in equation~\eqref{eq:coordinate.Asky} is also the position of astrometric origin measured from \sgra at $t\ls{S.apo}$. 
Secondly, because the velocities $\vec{V}\ls{astro}$ and $\vec{V}\ls{O}$ are, respectively, measured from C and \sgra, the relative velocity of the astrometric origin measured from \sgra is given by $\vec{V}\ls{astro}+\vec{V}_{XY}$, where $\vec{V}_{XY}$ is the projection of $\vec{V}\ls{O}$ onto the $X$-$Y$ plane.

\section{Derivation of $z\ls{1PN.0PM}(t)$ and some analyses of $\Delta z\ls{1PN.0PM}(t)$}
\label{app:approx}

\subsection{Derivation of the GR redshift~\eqref{eq:theory.z1PN0PM}}
\label{app:approx.derivation}

This appendix is for the derivation of the GR redshift of photons coming from S0-2 at the 1PN+0PM approximation~\eqref{eq:theory.z1PN0PM}. 
This redshift can be obtained by substituting the definition of frequency~\eqref{eq:theory.nu} into the definition of redshift~\eqref{eq:theory.z} under the 1PN+0PM approximation.

Before introducing the PN expansion, let us consider the situation that the S0-2 is regarded as a test particle moving in the Kerr spacetime of mass $M\ls{SgrA}$ and spin angular momentum $J\ls{SgrA}$. 
The components of metric tensor of Kerr spacetime, $g_{\mu\nu}$, are read from the line element,
\eqab
\label{eq:approx.metric}
 \diff{s^2} &=& g_{\mu\nu}\diff{x^\mu}\diff{x^\nu}
\\
\nonumber
 &=&
 \displaystyle
 - \frac{\Sigma D}{Z} \diff{t}^2
 + \frac{Z}{\Sigma} \sin^2\theta
    \left( \omega\, \diff{t} - \diff{\varphi} \right)^2
 + \frac{\Sigma}{D} \diff{r}^2
 + \Sigma \diff{\theta}^2
 \,,
\eqae
where the coordinates $x^\mu = (t,r,\theta,\varphi)$ are the Boyer-Lindquist system, and the functions in the metric components are
\eqab
 D(r) &=& r^2 + a^2 - 2 m\, r
\\
 \Sigma(r,\theta) &=& r^2 + a^2 \cos^2\theta
\\
 Z(r,\theta) &=& (r^2+a^2) \Sigma(r,\theta) + 2m\,r\, a^2 \sin^2\theta
\\
 \displaystyle
 \omega(r,\theta) &=& \frac{2 m\, a\, r}{Z(r,\theta)}
 \,,
\eqae
where the mass parameter $m = GM\ls{SgrA}/c^2$, the spin parameter $a = J\ls{SgrA}/(cM\ls{SgrA})$. 
These $m$ and $a$ have the dimension of length.

In the usual GR discussion, the spatial velocity of S0-2 is defined by using a tetrad basis. 
In our situation, it is reasonable to use the tetrad basis associated with the so-called ``locally non-rotating frame (LNRF)'' in Kerr spacetime. 
The unit timelike vector in the tetrad basis of LNRF, $e_{(t)}^\mu$, is perpendicular to the spacelike hypersurface at $t=$ constant in the outside of BH horizon,
\eqb
\label{eq:approx.LNRF}
 e_{(t)}^\mu =
 \left(\,
 \sqrt{\frac{Z}{\Sigma D}} \,,\, 0 \,,\, 0 \,,\, \omega\sqrt{\frac{Z}{\Sigma D}}
 \,\right)
 \,,
\eqe
where the components are given with the Boyer-Lindquist coordinates. 
As the spacelike unit vectors that compose the tetrad basis with $e_{(t)}^\mu$, we adopt the following three vectors,
\eqab
 \displaystyle
 e_{(r)}^\mu &=& \left(\, 0 \,,\, \sqrt{\frac{D}{\Sigma}} \,,\, 0 \,,\, 0 \,\right)
\\
 \displaystyle
 e_{(\theta)}^\mu &=& \left(\, 0 \,,\, 0 \,,\, \frac{1}{\sqrt{\Sigma}} \,,\, 0 \,\right)
\\
 \displaystyle
 e_{(\varphi)}^\mu &=&
 \left(\, 0 \,,\, 0 \,,\, 0 \,,\, \frac{1}{\sin\theta}\sqrt{\frac{\Sigma}{Z}} \,\right)
 \,,
\eqae
where the components are given with the Boyer-Lindquist coordinates. 
By definition of the tetrad basis, the orthonormal conditions are satisfied, $g_{\mu\nu}e_{(\alpha)}^\mu e_{(\beta)}^\nu = \eta_{(\alpha)(\beta)}$, where $\eta_{(t)(t)}=-1$, $\eta_{(t)(i)} = 0$, $\eta_{(i)(j)} = \delta_{(i)(j)}$ (Kronecker's delta), and $i ,j = r, \theta, \varphi$. 
Then, we define the spatial velocity of S0-2 in the context of GR, $\vec{V}\ls{S.GR}$, using the tetrad components of the four-velocity of S0-2, $U\ls{S}^\mu$,
\eqb
\label{eq:approx.VS}
 \displaystyle
 V\ls{S.GR}^i \defeq
 \frac{\phantom{-}g_{\mu\nu}e_{(i)}^\mu U\ls{S}^\nu}
        {-g_{\mu\nu}e_{(t)}^\mu U\ls{S}^\nu}
 \,,
\eqe
where $i = r, \theta, \varphi$, and all spacetime coordinates substituted into this formula are just at the spacetime position of S0-2, $x\ls{S}^\mu(\tau) = (t\ls{S}(\tau),r\ls{S}(\tau),\theta\ls{S}(\tau),\varphi\ls{S}(\tau)\,)$, that are the solution of the equations of motion~\eqref{eq:theory.geodesic.full}.

By the definition of LNRF, the time like vector $e_{(t)}^\mu$ has no angular velocity with respect to the spacelike hypersurface at $t=$ constant. 
However, this vector has a non-zero $\varphi$-component, $e\ls{(t)}^\varphi \neq 0$, in the Boyer-Lindquist coordinates. 
The angular velocity of $e_{(t)}^\mu$ measured in the Boyer-Lindquist coordinates (not in a coordinate system fixed to the hypersurface at $t=$ constant), $e_{(t)}^{\varphi}/e_{(t)}^t = \omega(r,\theta)$, is regarded as the angular velocity of the so-called ``frame dragging effect'' of a Kerr BH measured in the Boyer-Lindquist coordinates. 
However, as shown below, the frame dragging effect cannot be detected within the observational precision of current telescopes (that corresponds to the 1PN and 0PM approximations of GR).

Next, let us proceed to introduce the PN expansion. 
The small parameter of the PN expansion, $\varepsilon$, is given in equation~\eqref{eq:theory.PNparameter}. 
Using this $\varepsilon$, the components of the inverse metric, $g^{\mu\nu}$, are expanded to be
\eqab
 g^{tt} &=&
 -1 - \frac{2m}{r}\bar{\varepsilon} - \frac{4m^2}{r^2}\bar{\varepsilon}^2
\nonumber
\\
 &&
 -\frac{2m (4m^2 - a^2 \cos^2\theta)}{r^3}\bar{\varepsilon^3} + O(\varepsilon^4)
\\
 g^{t\varphi} &=&
 -\frac{2m\,a}{r^3}\bar{\varepsilon}^3 + O(\varepsilon^4)
\\
 g^{rr} &=&
 1 - \frac{2m}{r}\bar{\varepsilon} + \frac{a^2}{r^2}\sin^2\theta\,\bar{\varepsilon}^2
 +\frac{2m\,a^2}{r^3}\bar{\varepsilon^3} + O(\varepsilon^4)
\\
 g^{\theta\theta} &=&
 \frac{1}{r^2}\bar{\varepsilon^2} + O(\varepsilon^4)
\\
 g^{\varphi\varphi} &=&
 \frac{1}{r^2 \sin^2\theta}\bar{\varepsilon^2} + O(\varepsilon^4)
 \,,
\eqae
where an auxiliary parameter $\bar{\varepsilon}$ is introduced to count the order of the PN expansion, for example the term $4m^2\bar{\varepsilon}^2/r^2$ is understood as the 2PN order term. 
Although we need only the 1PN approximation within the present observational precision, as discussed in section~\ref{sec:theory.approximation}, the PN expansion of $g^{\mu\nu}$ up to some higher order terms shown in the above equations may be useful for readers who will follow our theoretical calculations, because those higher order terms are necessary to obtain the appropriate form of the Hamiltonian of S0-2 at the 1PN approximation~\eqref{eq:theory.H.1PN}. 
On the other hand, for our purpose, it is enough to expand the LNRF tetrad basis up to the 1PN order,
\eqab
 e_{(t)}^\mu &=&
 \left(\, 1+ \frac{m}{r}\bar{\varepsilon} + O(\varepsilon^2)
 \,,\, 0 \,,\, 0 \,,\, O(\varepsilon^3) \,\right)
\\
 e_{(r)}^\mu &=&
 \left(\, 0 \,,\, 1- \frac{m}{r}\bar{\varepsilon} + O(\varepsilon^2) \,,\, 0 \,,\, 0 \,\right)
\\
 e_{(\theta)}^\mu &=&
 \left(\, 0 \,,\, 0 \,,\, 1- \frac{1}{r}\bar{\varepsilon} + O(\varepsilon^2) \,,\, 0 \,\right)
\\
 e_{(\varphi)}^\mu &=&
 \left(\, 0 \,,\, 0 \,,\, 0 \,,\, 1- \frac{1}{r \sin\theta}\bar{\varepsilon} + O(\varepsilon^2) \,\right)
 \,.
\eqae
Note that, in the expansions of $g^{\mu\nu}$ and $e_{(\nu)}^\mu$, the auxiliary parameter $\bar{\varepsilon}$ should be set at unity, $\bar{\varepsilon} = 1$, after finishing the calculation of the PN expansion, because $\bar{\varepsilon}$ is simply introduced in order to count the order of the PN expansion.

The four-velocity of S0-2, $U\ls{S}^\mu$, at the 1PN approximation is obtained by substituting the 1PN form of the LNRF tetrad into $\vec{V}\ls{S.GR}$ defined in equation~\eqref{eq:approx.VS}. 
In this calculation, we need to take two items into account; (i) the normalization condition, $g_{\mu\nu}U\ls{S}^\mu U\ls{S}^\nu = -1$, and (ii) the relation, $(\vec{V}\ls{S.GR}/c)^2 \sim m/r\ls{s.1PN} \approx \varepsilon$, implied by the fact that S0-2 is gravitationally bounded by \sgra. 
Then, we obtain,
\eqb
\label{eq:approx.US}
 U\ls{S}^\mu =
 \left(\, 1 + \frac{1}{2}\Bigl(\frac{\vec{V}\ls{G.1PN}}{c}\Bigr)^2 + \frac{m}{r\ls{S.1PN}} \,,\,
 \vec{V}\ls{S.1PN} \,\right) \,,
\eqe
where $\vec{V}\ls{S.1PN}$ is the spatial velocity of S0-2 at the 1PN approximation, and $r\ls{S.1PN}(\tau)$ is the radial coordinate of S0-2 that is the solution of geodesic equations at the 1PN approximation. 
We find, at the 1PN approximation, the spatial components of $U\ls{S}^\mu$ are equal to $\vec{V}\ls{S.1PN}$.

Remember that, as discussed in section~\ref{sec:theory.approximation}, we assume that the spatial velocity of the observer, $\vec{V}\ls{O}$, is constant and free from the gravity of \sgra. 
This indicates that the special relativistic form is applicable to the four-velocity of the observer, 
\eqb
\label{eq:approx.UO}
 U\ls{O}^\mu =
 \left(\, \gamma\ls{O} \,,\, \gamma\ls{O} \vec{V}\ls{O} \,\right)
 \,,
\eqe
where $\gamma\ls{O} = 1/\sqrt{1-(\vec{V}\ls{O}/c)^2}$.

Next, let us introduce the 0PM approximation of the null vector tangent to the null geodesics of photons coming from S0-2 to the observer,
\eqb
 K^\mu = \left(\, K^t \,,\, \vec{K} \,\right) \,.
\eqe
This is the four-wave-vector of the photon. 
At the 0PM approximation, the null geodesic is approximated as a straight line connecting the emission event of the photon by S0-2 and the observation event of the photon by the observer. 
Within this approximation, the spatial directional vector at the emission, $\vec{K}\ls{S.0PM}$, and the vector at the observation, $\vec{K}\ls{O.0PM}$, are parallel,
\eqb
 \vec{K}\ls{S.0PM} \propto \vec{K}\ls{O.0PM} \,.
\eqe
However, because the dispersion relation of the photon is given by the null condition, $g^{\mu\nu}K_\mu K_\nu = 0$, the frequency of the photon varies according to the spacetime position on the straight null geodesic. 
At the emission event, the null condition gives,
\eqb
 \bigl|\vec{K}\ls{S.1PN0PM}\bigr| = \left(\, 1+\frac{2m}{r\ls{S.1PN}} \,\right) F \,,
\eqe
where the 1PN approximation of the metric tensor is used, and $F = - K_t = g_{t\mu}K^\mu$ is a constant conserved along the null geodesic due to the stationarity of BH spacetime. 
Then, using the same constant $F$, the null condition at the observation event gives,
\eqb
 \bigl|\vec{K}\ls{O.1PN0PM}\bigr| = F \,.
\eqe

Finally, we collect the above preparations in order to calculate the GR redshift at the 1PN+0PM approximation. 
By substituting the above 1PN+0PM formulas of $g_{\mu\nu}$, $U\ls{S}^\mu$, $U\ls{O}^\mu$ and $K^\mu$ into the definition of frequency~\eqref{eq:theory.nu}, we obtain the frequency at the emission event,
\eqb
 \nu\ls{S.1PN0PM} =
 \left(\, 1 + \frac{m}{r\ls{S.1PN}} \,\right)
 \left(\, 1 - V\ls{S.1PN}\us{(K)} + \frac{\vec{V}\ls{S.1PN}^2}{2}
 \,\right) F
 \,,
\eqe
and the frequency at the observation event,
\eqb
 \nu\ls{O.1PN0PM} =
 \left(\, 1 + \frac{1}{2}\Bigl(\frac{\vec{V}\ls{O}}{c}\Bigr)^2 \,\right)
 \left(\, 1 - V\ls{O.1PN}\us{(K)} \,\right) F
 \,,
\eqe
where $V\ls{S.1PN}\us{(K)}$ (and $V\ls{O.1PN}\us{(K)}$) is the component of $\vec{V}\ls{S.1PN}$ (and $\vec{V}\ls{O.1PN}$) that is parallel to $\vec{K}\ls{S.1PN0PM}$, whose positive direction is from S0-2 to the observer. 
Here note that, the ``line-of-sight'' direction introduced in section~\ref{sec:theory.definition} is parallel to $\vec{K}\ls{S.1PN0PM}$ but the positive direction is opposite, $V\ls{S.1PN}\us{(K)} = -V\ls{S.1PN \parallel}$ and $V\ls{O.1PN}\us{(K)} = -V\ls{O.1PN \parallel}$. 
Hence, by substituting the above frequencies into the definition of the GR redshift~\eqref{eq:theory.z}, we finally obtain the redshift at the 1PN+0PM approximation, $z\ls{1PN.0PM}(t)$ in equation~\eqref{eq:theory.z1PN0PM}. 
Further note that, because the spin parameter, $a$, does not appear in equation~\eqref{eq:theory.z1PN0PM}, the component of GR effect depending on the BH spin cannot be observed within the present observational precision.

\subsection{Some analyses of the GR evidence~\eqref{eq:theory.dz1PN0PM}}
\label{app:approx.analyses}

The GR evidence under our presupposition on the parameter values, $\Delta z\ls{1PN.0PM}(t)$ at the 1PN+0PM approximation, is given in the equation~\eqref{eq:theory.dz1PN0PM}. 
For a deeper understanding of $\Delta z\ls{1PN.0PM}(t)$, let us make some theoretical analyses. 
The temporal component of the geodesic equations at the 1PN approximation reads
\eqb
\label{eq:approx.t1PN}
 \od{t\ls{GR}(\tau)}{\tau} = 
 1 + \frac{2GM\ls{SgrA}}{c^2 r\ls{S.1PN}(\tau)} \,,
\eqe
where $t\ls{GR}$ is the coordinate time in GR (not in the Newtonian case), and $\tau$ is the proper time of S0-2. 
From this equation, we can estimate as
\eqb
\label{eq:approx.tGR.tau}
 t\ls{GR}
 \,\approx\, \tau + \frac{2GM\ls{SgrA}}{c^2 r\ls{S.1PN}} \delta\tau
 \,\approx\, \tau + \varepsilon \delta\tau \,,
\eqe
where $\varepsilon$ is the PN parameter~\eqref{eq:theory.PNparameter}, and the order of $\delta\tau$ may be estimated by a typical time scale of our system,
\eqb
 \delta\tau \sim O(r\ls{peri}/c) \,,
\eqe
where $r\ls{peri}$ is the pericenter distance of S0-2 to \sgra.

In comparing the GR prediction with the Newtonian prediction, one may consider that the Newtonian time $t\ls{NG}$ corresponds to the proper time of S0-2, $t\ls{NG} \leftrightarrow \tau$, or to the Lorentz-transformed case, $t\ls{NG} \leftrightarrow \tau/\sqrt{1-(V\ls{S}/c)^2}$. 
However, the difference between these correspondences of the temporal coordinates do not affect the following discussions at the 1PN+0PM approximation. 
Note that the latter correspondence is estimated as, $t\ls{NG} \leftrightarrow \tau/\sqrt{1-(V\ls{S}/c)^2} \approx (1 + \varepsilon ) \tau$, where the order relation $V\ls{S}/c \sim O(\varepsilon^{1/2})$ is used. 
The term $\varepsilon\tau$ can be absorbed into the second term in equation~\eqref{eq:approx.tGR.tau}. 
Therefore, the latter correspondence degenerates to the former one, $t\ls{NG} \leftrightarrow \tau$, at the 1PN+0PM approximation. 
This correspondence of the temporal coordinates is assumed in the following discussions.

The position of S0-2 may be expanded as
\eqb
\label{eq:approx.xS1PN}
 \vec{x}\ls{S}(t\ls{GR}) \approx
 \vec{x}\ls{S.1PN}(\tau)
 + \varepsilon\vec{V}\ls{S.1PN}(\tau)\, \delta\tau \,.
\eqe
The second term is of the 1.5PN order because of the order relation  $V\ls{S}/c \sim O(\varepsilon^{1/2})$. 
Therefore, the Roemer time delay equation~\eqref{eq:theory.tR} at the 1PN+0PM approximation is determined by the first term of equation~\eqref{eq:approx.xS1PN} within the present observational precision. 
Note that, one may count the first term, $\vec{x}\ls{S.1PN}(\tau)$, as a 0PN approximation term, but the parameter values in this term is determined by fitting the given observational data with the 1PN+0PM motion of S0-2 under our presupposition on the parameter values. 
This denotes that the Roemer time delay in the GR redshift does not necessarily equal the one in the Newtonian redshift, $t\ls{R.1PN0PM} \neq t\ls{R.NG}$, under our presupposition.

Next, the spatial velocity of S0-2 may be expanded as
\eqb
\label{eq:approx.VS1PN}
 \vec{V}\ls{S}(t\ls{GR}) \approx
 \vec{V}\ls{S.1PN}(\tau)
 +\delta\tau \dot{\vec{V}}\ls{S.1PN}(\tau) \,,
\eqe
where $\dot{\vec{V}}\ls{S.1PN}$ is the acceleration of S0-2. 
Note that one may count the first term in equation~\eqref{eq:approx.VS1PN} as a 0PN approximation term, but the parameter values in this term is the best-fitting values in the 1PN+0PM approximation under our presupposition on the parameter values. 
Furthermore, the second term in equation~\eqref{eq:approx.VS1PN} is of the 1PN order, because the term is estimated by the equations of motion as
\eqb
 \delta\tau \dot{\vec{V}}\ls{S.1PN} 
 \,\approx\,
 \delta\tau \frac{GM\ls{SgrA}}{r\ls{peri}^2}
 \,\approx\,
 c\varepsilon \,.
\eqe
Therefore, the spatial velocity of S0-2 in the GR redshift at the 1PN+0PM approximation is different from the one in the Newtonian redshift, $\vec{V}\ls{S.1PN}(t) \neq \vec{V}\ls{S.NG}(t)$. 
This result, together with the result in the previous paragraph on the Roemer time delay, denote that the first term in the GR evidence~\eqref{eq:theory.dz1PN0PM}, $[V\ls{S.1PN\parallel}(t+t\ls{R}) - V\ls{S.NG\parallel}(t+t\ls{R})]/c$, does not vanish and has to be counted as a non-vanishing component in $\Delta z\ls{1PN.0PM}(t)$.

The squared velocity of S0-2 is estimated as
\eqb
\label{eq:approx.VS1PN2}
 \vec{V}\ls{S}(t\ls{GR})^2 \approx
 \vec{V}\ls{S.1PN}(\tau)^2 + c^2 \varepsilon^{1.5} \,,
\eqe
where the order relation, $V\ls{S}/c \sim O(\varepsilon^{1/2})$, is used. 
Because the second term is of the 1.5PN order, the squared velocity at the 1PN+0PM approximation is actually determined by the first term of equation~\eqref{eq:approx.VS1PN2}. 
Note that, one may count the first term as a 0PN approximation term, but the parameter values in this term is the best-fitting values in the 1PN+0PM approximation under our presupposition on the parameter values. 
This fact, together with the result on the Roemer time delay, denote that the third term in the GR evidence~\eqref{eq:theory.dz1PN0PM}, $[\vec{V}\ls{S.1PN}(t+t\ls{R})^2 - \vec{V}\ls{O.1PN}(t)^2]/(2c^2)$, under our presupposition is not a purely special relativistic term. 
The reason is as follows: 
If one wants to calculate the purely special relativistic value of this term, then the motion of S0-2 has to be Newtonian, because the special relativity is the theoretical framework that ignores the effect of the spacetime curvature (which is the GR's own gravitational effect and does never arises in the framework of special relativity). 
Hence, because the parameter values used in the third term of $\Delta z\ls{1PN.0PM}$ is not the Newtonian values but the 1PN+0PM values under our presupposition on the parameter values, the resultant value of the third term of $\Delta z\ls{1PN.0PM}$ has to be interpreted as a non-linear combination of the special relativistic and GR predictions.


\end{document}